# Identifying Young Stars in Massive Star-Forming Regions for the MYStIX Project

Patrick S. Broos[1], Konstantin V. Getman[1], Matthew S. Povich[1,2], Eric D. Feigelson[1], Leisa K. Townsley[1]

Tim Naylor[3], Michael A. Kuhn[1], R. R. King[3], Heather A. Busk[1]

patb@astro.psu.edu

## ABSTRACT

The **M**assive **Y**oung star-forming Complex **St**udy in **I**nfrared and **X**-rays (MYStIX) project requires samples of young stars that are likely members of 20 nearby Galactic massive star-forming regions. Membership is inferred from statistical classification of X-ray sources, from detection of a robust infrared excess that is best explained by circumstellar dust in a disk or infalling envelope, and from published spectral types that are unlikely to be found among field stars. We present the MYStIX membership lists here, and describe in detail the statistical classification of X-ray sources via a "Naive Bayes Classifier." These membership lists provide the empirical foundation for later MYStIX science studies.

*Subject headings:* methods: data analysis — methods: statistical — stars: pre-main sequence — X-rays: general — X-rays: stars — infrared: stars — open clusters and associations: general

## 1. Introduction

The **M**assive **Y**oung star-forming Complex **St**udy in **I**nfrared and **X**-rays (MYStIX) project, described by Feigelson et al. (2013), seeks to identify and study samples of young stars in 20 nearby ($0.4 < d < 3.6$ kpc) Galactic massive star-forming regions (MSFRs). These samples are derived using X-ray data from the *Chandra X-ray Observatory*, near-infrared (NIR) photometry from the United Kingdom InfraRed Telescope (UKIRT, Casali et al. 2007) and from 2MASS, mid-infrared (MIR) photometry from the *Spitzer Space Telescope*, and published spectroscopically identified massive stars. Membership in an MSFR is almost always uncertain due to several sources of contaminants. Our purpose here is to describe our efforts to minimize contaminants in the MYStIX catalogs of young stars, which we refer to as "MYStIX Probable Complex Members" (MPCMs), and to present the MPCM catalog for each MYStIX MSFR. These catalogs will be used in astronomical and astrophysical studies of the young stellar populations in these regions.

An MPCM catalog is the union of three sets of probable members identified by three established methods for identifying young stars (Feigelson et al. 2013, Figure 3). The majority of members (∼1000 per MSFR) are identified via a statistical classification of X-ray point sources similar to that developed for the *Chandra* Carina Complex Project (CCCP, Townsley et al. 2011) by Broos et al. (2011b). The MYStIX X-ray source classification procedure is described in Section 3. Hundreds of additional members of each

[1]Department of Astronomy & Astrophysics, 525 Davey Laboratory, Pennsylvania State University, University Park, PA 16802, USA

[2]Department of Physics and Astronomy, California State Polytechnic University, 3801 West Temple Ave, Pomona, CA 91768

[3]School of Physics and Astronomy, University of Exeter, Stocker Road, Exeter, EX4 4QL, UK

MSFR, not detected by *Chandra*, are identified by modeling their NIR/MIR spectral energy distributions (SEDs) and then adopting as members those objects with a robust infrared excess that is best explained by circumstellar dust in a disk or infalling envelope (Povich et al. 2011). Details of our SED modeling and lists of infrared excess sources over fields of view wider than the MYStIX fields are presented by Povich et al. (2013). Stars with spectral types B3 or earlier are also added to the MPCM catalog, based on the assumption that massive stars are unlikely to lie in the foreground or background.

For the convenience of readers who are most interested in the results of our membership studies, this paper first presents the MYStIX MPCM catalogs (Section 2, Table 2). Those wide tables collate a large number of X-ray and infrared source properties published elsewhere, so that the MPCM catalogs are immediately useful to the reader without cross-referencing among X-ray tables, infrared tables, and catalog matching tables. Readers who are interested in the details of our X-ray classification procedure, or are interested in an electronic table containing the classification results for *all* MYStIX X-ray sources, should carry on with Sections 3—6 and with Appendix A. Finally, Section 7 contrasts the classification method presented here with those historically applied to star-forming regions.

## 2. The MPCM Catalogs

For each MYStIX region, Table 1 presents source tallies from the X-ray/NIR/MIR observations, tallies of matches between X-ray and IR catalogs, tallies of the X-ray source classifications (described in Section 3), and finally the number of sources in the MPCM catalog.

Table 2 defines the columns of an electronic MPCM catalog that is available in ASCII format from the electronic edition of this article and that may be available in many other formats from Vizier (Ochsenbein et al. 2000). For the reader's convenience, the MPCM catalog reproduces X-ray properties presented by Townsley et al. (2013) and by Kuhn et al. (2013a). Some columns characterize the extracted X-ray spectrum (e.g., *MedianEnergy*); some characterize the apparent spectrum incident on *Chandra* (e.g., log *PhotonFlux*), and some characterize the astrophysical spectrum corrected for interstellar absorption (e.g., *LOGNH_OUT* and *LX*).[1] Townsley et al. (2013) and Kuhn et al. (2013a) identify a few very bright X-ray sources in each region that suffer from a type of instrumental non-linearity known as *photon pile-up*[2]; X-ray properties reported in Table 2 for those sources are biased and should not be used for quantitative work.

The MPCM catalog also reproduces infrared photometry presented by other MYStIX papers. NIR catalogs were constructed by combining deep UKIRT catalogs (King et al. 2013) where available, with bright stars from the 2MASS catalog (Cutri et al. 2003). MIR catalogs were obtained from local reductions of *Spitzer* observations (Kuhn et al. 2013b), from the *Spitzer* Galactic Legacy Infrared Mid-Plane Survey Extraordinaire (GLIMPSE; Benjamin et al. 2003), and from the Vela-Carina Survey (*Spitzer* Proposal ID 40791, PI S. Majewski). Feigelson et al. (2013, Table 2) report which IR catalogs were available for each MYStIX MSFR. Potential NIR and MIR counterparts to X-ray sources were identified by the statistical catalog matching method described by Naylor et al. (2013), which estimates the probability that each NIR and MIR source is the counterpart to each X-ray source. We report an IR counterpart when its counterpart probability is larger than 0.80.

---

[1] Intrinsic (absorption-corrected) X-ray luminosity is estimated by the *XPHOT* algorithm described by Getman et al. (2010), under the assumption that the object is a low-mass pre-main sequence star at the distance assumed by MYStIX (Feigelson et al. 2013, Table 1). When a source has been associated with a known massive star (column OB_LABEL) all XPHOT quantities will be less reliable. When a non-member source has been mistakenly included in the MPCM catalog, our distance assumption and the luminosity estimate will be wrong.

[2] http://cxc.harvard.edu/ciao/why/pileup_intro.html

MSFR, not detected by *Chandra*, are identified by modeling their NIR/MIR spectral energy distributions BAD (Wenger et al. 2000). Spectral types are mostly based on visual-band spectra obtained by many researchers, but occasionally are based on near-infrared or ultraviolet spectroscopy. In the Carina Nebula and Orion Nebula, the OB lists were obtained from Gagné et al. (2011) and Stelzer et al. (2005), respectively. Note that these spectroscopically identified OB stars typically do not have IR excesses and some, but not all, have X-ray detections. Since the published positions of some OB stars have significant uncertainty, direct matching between the X-ray and OB catalogs is not appropriate. Instead, each OB star was identified in our NIR catalog, and if that NIR source is declared as an X-ray counterpart then we also declare the OB star to be detected by *Chandra*.

The MYStIX source position (Class_RAdeg,Class_DEdeg) reported in Table 2 is the most accurate position among the multiwavelength detections that we judged to be the same object—a *Chandra* X-ray position, UKIRT or 2MASS NIR position, or a *Spitzer* MIR position. The MYStIX source name (Class_Name) is the sexagesimal representation of the MYStIX source position. In rare cases two MPCM rows will share the same source position, because two X-ray sources have been assigned the same NIR counterpart by our counterpart identification algorithm. In such cases we append "a" or "b" to the source names to make them unique.

Table 1. Source Tallies in MYStIX MSFRs[a]

| Line | Population | Orion Nebula | Flame Nebula | W 40 | RCW 36 | NGC 2264 | Rosette Nebula | Lagoon Nebula | NGC 2362 | DR 21 | RCW 38 |
|---|---|---|---|---|---|---|---|---|---|---|---|
| | Single-wavelength results | | | | | | | | | | |
| 1 | *Chandra* X-ray sources | 1616 | 547 | 225 | 502 | 1328 | 1962 | 2427 | 690 | 765 | 1019 |
| 2 | UKIDSS/2MASS NIR sources | ... | 754 | 2255 | 1446 | 11865 | 37816 | 90772 | 7887 | 22142 | 2737 |
| 3 | *Spitzer* MIR sources | ... | 4019 | 14120 | 1632 | 10284 | 14383 | 31534 | 8261 | 15923 | 2499 |
| 4 | Published OB stars | 13 | 2 | 3 | 0 | 8 | 23 | 28 | 12 | 1 | 3 |
| | Multi-wavelength results | | | | | | | | | | |
| 5 | X-ray/NIR matches [b] | ... | 261 | 185 | 254 | 753 | 1246 | 1461 | 447 | 407 | 409 |
| 6 | X-ray/MIR matches [b] | ... | 292 | 184 | 172 | 769 | 1200 | 1011 | 488 | 361 | 309 |
| 7 | X-ray/(NIR or MIR) matches [b] | ... | 302 | 191 | 267 | 799 | 1300 | 1483 | 503 | 447 | 450 |
| 8 | NIR/MIR SED excess sources | 631 | 193 | 308 | 135 | 556 | 622 | 468 | 67 | 507 | 112 |
| | X-ray detected | 521 | 131 | 78 | 88 | 282 | 237 | 253 | 29 | 122 | 39 |
| | X-ray undetected | 110 | 62 | 230 | 47 | 274 | 385 | 215 | 38 | 385 | 73 |
| | X-ray source classification results | | | | | | | | | | |
| 9 | X-ray foreground stars[c] | ... | 0 | 10 | 9 | 0 | 4 | 2 | 0 | 4 | 13 |
| 10 | X-ray background stars[c] | ... | 0 | 0 | 0 | 0 | 0 | 3 | 0 | 5 | 1 |
| 11 | X-ray extragalactic objects[c] | ... | 7 | 19 | 0 | 126 | 190 | 102 | 119 | 0 | 1 |
| 12 | X-ray young stars[c] | 1414 | 422 | 194 | 337 | 898 | 1337 | 1828 | 467 | 594 | 813 |
| 13 | X-ray unclassified[c] | ... | 118 | 2 | 156 | 304 | 431 | 492 | 104 | 162 | 191 |
| | Young star catalog | | | | | | | | | | |
| 14 | MYStIX Probable Complex Members | 1524 | 485 | 426 | 384 | 1173 | 1731 | 2056 | 510 | 980 | 886 |

Table 1. (continued) Source Tallies in MYStIX MSFRs[a]

| Line | Population | NGC 6334 | NGC 6357 | Eagle Nebula | M 17 | W 3 | W 4 | Carina Nebula[d] | Trifid Nebula | NGC 3576 | NGC 1893 |
|---|---|---|---|---|---|---|---|---|---|---|---|
| | Single-wavelength results | | | | | | | | | | |
| 1 | *Chandra* X-ray sources | 1510 | 2360 | 2830 | 2999 | 2094 | 647 | 7412 | 633 | 1522 | 1442 |
| 2 | UKIDSS/2MASS NIR sources | 136283 | 207319 | 200331 | 224019 | 6751 | 2781 | ... | 11865 | 12737 | 10625 |
| 3 | *Spitzer* MIR sources | 28158 | 45878 | 43126 | 45227 | 9900 | 10296 | ... | 10284 | 12732 | 9414 |
| 4 | Published OB stars | 8 | 17 | 67 | 64 | 23 | 37 | 134 | 2 | 11 | 34 |
| | Multi-wavelength results | | | | | | | | | | |
| 5 | X-ray/NIR matches [b] | 1063 | 1649 | 1687 | 1861 | 927 | 334 | 6367 | 355 | 617 | 965 |
| 6 | X-ray/MIR matches [b] | 568 | 1159 | 1250 | 738 | 738 | 412 | 3831 | 240 | 525 | 943 |
| 7 | X-ray/(NIR or MIR) matches [b] | 1082 | 1730 | 1742 | 1906 | 1038 | 415 | 6474 | 364 | 677 | 1053 |
| 8 | NIR/MIR SED excess sources | 407 | 523 | 721 | 155 | 259 | 155 | 815 | 174 | 142 | 538 |
| | X-ray detected | 127 | 243 | 239 | 110 | 164 | 66 | 283 | 60 | 66 | 173 |
| | X-ray undetected | 280 | 280 | 482 | 45 | 95 | 89 | 532 | 114 | 76 | 365 |
| | X-ray source classification results | | | | | | | | | | |
| 9 | X-ray foreground stars[c] | 8 | 11 | 7 | 73 | 36 | 3 | 160 | 3 | 1 | 7 |
| 10 | X-ray background stars[c] | 0 | 0 | 1 | 107 | 3 | 0 | 0 | 10 | 0 | 0 |
| 11 | X-ray extragalactic objects[c] | 2 | 13 | 106 | 47 | 29 | 71 | 104 | 38 | 0 | 132 |
| 12 | X-ray young stars[c] | 1385 | 1952 | 2065 | 2296 | 1571 | 411 | 6751 | 418 | 1131 | 1110 |
| 13 | X-ray unclassified[c] | 115 | 384 | 651 | 476 | 455 | 162 | 397 | 164 | 390 | 193 |
| | Young star catalog | | | | | | | | | | |
| 14 | MYStIX Probable Complex Members | 1668 | 2240 | 2582 | 2365 | 1676 | 522 | 7334 | 533 | 1214 | 1495 |

[a] Spatially restricted to X-ray field of view
[b] Counterpart probability >0.80 (Naylor et al. 2013).
[c] Includes X-ray sources only
[d] Restricted to the field of view of the HAWK-I NIR observations (Preibisch et al. 2011).

Table 2. MPCM Sources and Properties

| Short Label (1) | Long Label (2) | Units (3) | Description (4) |
|---|---|---|---|

[The short labels in Column 1 were composed by the ApJ for the CCCP electronic table; they should be re-used in the MYStIX electronic table. The long labels in Column 2 **without parentheses** should be printed here to link these columns to Broos et al. (2011a, Table 1, print version). The long labels in Column 2 **with parentheses** have not appeared in the literature and should be removed after Col. 1 is filled in with short labels.

**MYStIX coordinates** (Section 2)

| | | | |
|---|---|---|---|
| TBD | (MYSTIX_SFR) | ⋯ | MSFR name |
| TBD | (Class_Name) | ⋯ | IAU source name; prefix is **J** |
| TBD | (Class_RAdeg) | deg | right ascension (J2000) |
| TBD | (Class_DEdeg) | deg | declination (J2000) |
| TBD | (Class_Pos_Err) | arcsec | 1-$\sigma$ error circle around (RAdeg,DEdeg) |
| TBD | (Class_Pos_Origin) | | origin of position |

**Multi-wavelength Detections**

| | | | |
|---|---|---|---|
| TBD | (Xray_Name) | ⋯ | X-ray source name in IAU format |
| TBD | (Xray_Label[a]) | ⋯ | X-ray source name used within the MYStIX project |
| TBD | (NIR_NAME) | | name in NIR catalog |
| TBD | (NIR_LABEL) | | label in NIR catalog |
| TBD | (MIR_NAME) | | name in MIR catalog |
| TBD | (MIR_LABEL) | | label in MIR catalog |
| TBD | (OB_LABEL) | | label in OB catalog |
| TBD | (XCAT_INDEX) | | 0-based index in X-ray catalog (Section 4) |
| TBD | (ISED_INDEX) | | 0-based index in IR excess catalog (Povich et al. 2013) |

**OB Properties** (from Skiff 2009; Wenger et al. 2000)

| | | | |
|---|---|---|---|
| TBD | (SPTY) | | spectral type |
| TBD | (ORIGIN_OB) | | reference for spectral type |
| TBD | (MAG_OB) | mag | visual photometry |
| TBD | (BAND_OB) | | visual band for MAG_OB |

**X-ray Observation**

| | | | |
|---|---|---|---|
| PNoSrc-m | ProbNoSrc_min | ⋯ | p-value[b] for no-source hypothesis (Broos et al. 2010, Section 4.3) |
| PKS-s | ProbKS_single[c] | ⋯ | smallest p-value for the one-sample Kolmogorov-Smirnov statistic under the no-variability null hypothesis within a single-observation |
| PKS-m | ProbKS_merge[c] | ⋯ | smallest p-value for the one-sample Kolmogorov-Smirnov statistic under the no-variability null hypothesis over merged observations |
| ExpNom | ExposureTimeNominal | s | total exposure time in merged observations |
| ExpFrac | ExposureFraction[d] | ⋯ | fraction of ExposureTimeNominal that source was observed |
| NObs | NumObservations | ⋯ | total number of observations extracted |
| NMerge | NumMerged | ⋯ | number of observations merged to estimate photometry properties |
| e_Theta | Theta_Lo | arcmin | smallest off-axis angle for merged observations |
| Theta | Theta | arcmin | average off-axis angle for merged observations |
| E_Theta | Theta_Hi | arcmin | largest off-axis angle for merged observations |
| PSFFrac | PsfFraction | ⋯ | average PSF fraction (at 1.5 keV) for merged observations |
| AGlow | AfterglowFraction[e] | ⋯ | suspected afterglow fraction |
| SrCnt-t | SrcCounts_t | count | extracted counts in merged apertures |
| NCt-t | NetCounts_t | count | net counts in merged apertures |
| NCt-s | NetCounts_s | count | net counts in merged apertures |
| NCt-h | NetCounts_h | count | net counts in merged apertures |
| loNCt-t | NetCounts_Lo_t[f] | count | 1-sigma lower bound on NetCounts_t |
| upNCt-t | NetCounts_Hi_t | count | 1-sigma upper bound on NetCounts_t |

Table 2—Continued

| Short Label (1) | Long Label (2) | Units (3) | Description (4) |
|---|---|---|---|
| loNCt-s | NetCounts_Lo_s | count | 1-sigma lower bound on NetCounts_s |
| upNCt-s | NetCounts_Hi_s | count | 1-sigma upper bound on NetCounts_s |
| loNCt-h | NetCounts_Lo_h | count | 1-sigma lower bound on NetCounts_h |
| upNCt-h | NetCounts_Hi_h | count | 1-sigma upper bound on NetCounts_h |
| | | | |
| Eng-t | MedianEnergy_t[g] | keV | median energy, observed spectrum |
| | | | |
| logF | log_PhotonFlux_t[h] | photon /cm**2 /s | incident photon flux |
| TBD | log_PhotonFlux_s | photon /cm**2 /s | incident photon flux |
| TBD | log_PhotonFlux_h | photon /cm**2 /s | incident photon flux |
| **X-ray Spectral Model** (Getman et al. 2010, XPHOT) | | | |
| TBD | (LX_H) | erg /s | X-ray luminosity, 2:8 keV |
| TBD | (LX_HC) | erg /s | absorption-corrected X-ray luminosity, 2:8 keV |
| TBD | (SLX_HC_STAT) | erg /s | 1-sigma statistical uncertainty on LX_HC |
| TBD | (SLX_HC_SYST) | erg /s | 1-sigma systematic uncertainty on LX_HC |
| TBD | (LX_T) | erg /s | X-ray luminosity, 0.5:8 keV |
| logL | (LX_TC) | erg /s | absorption-corrected X-ray luminosity, 0.5:8 kev |
| TBD | (SLX_TC_STAT) | erg /s | 1-sigma statistical uncertainty on LX_TC |
| TBD | (SLX_TC_SYST) | erg /s | 1-sigma systematic uncertainty on LX_TC |
| logNH | (LOGNH_OUT) | /cm**2 | gas column density |
| TBD | (SLOGNH_OUT_STAT_OUT) | /cm**2 | 1-sigma statistical uncertainty on LOGNH_OUT |
| TBD | (SLOGNH_OUT_SYST_OUT) | /cm**2 | 1-sigma systematic uncertainty on LOGNH_OUT |
| **IR Counterparts and Photometry** | | | |
| TBD | (XN_PROB_CP) | | counterpart probability, X-ray/NIR (Naylor et al. 2013) |
| TBD | (XM_PROB_CP) | | counterpart probability, X-ray/MIR (Naylor et al. 2013) |
| TBD | (MAG_J) | mag | photometry |
| TBD | (ERROR_J) | mag | 1-sigma uncertainty |
| TBD | (MAG_H) | mag | photometry |
| TBD | (ERROR_H) | mag | 1-sigma uncertainty |
| TBD | (MAG_K) | mag | photometry |
| TBD | (ERROR_K) | mag | 1-sigma uncertainty |
| TBD | (MAG_3p6um) | mag | photometry |
| TBD | (ERROR_3p6um) | mag | 1-sigma uncertainty |
| TBD | (MAG_4p5um) | mag | photometry |
| TBD | (ERROR_4p5um) | mag | 1-sigma uncertainty |
| TBD | (MAG_5p8um) | mag | photometry |
| TBD | (ERROR_5p8um) | mag | 1-sigma uncertainty |
| TBD | (MAG_8p0um) | mag | photometry |
| TBD | (ERROR_8p0um) | mag | 1-sigma uncertainty |
| TBD | (J_FLAG) | | UKIRT photometry flag (King et al. 2013) |
| TBD | (H_FLAG) | | UKIRT photometry flag (King et al. 2013) |
| TBD | (K_FLAG) | | UKIRT photometry flag (King et al. 2013) |
| TBD | (CC_FLG) | | 2MASS photometry flag |
| TBD | (PH_QUAL) | | 2MASS photometry flag |
| TBD | (SQF_J) | | GLIMPSE photometry flag |
| TBD | (SQF_H) | | GLIMPSE photometry flag |
| TBD | (SQF_K) | | GLIMPSE photometry flag |
| TBD | (SQF_3P6UM) | | GLIMPSE photometry flag |
| TBD | (SQF_4P5UM) | | GLIMPSE photometry flag |
| TBD | (SQF_4P8UM) | | GLIMPSE photometry flag |
| TBD | (SQF_8P0UM) | | GLIMPSE photometry flag |
| TBD | (AP_LS_FLG) | | "Local Spitzer" photometry flag (Kuhn et al. 2013b) |

Table 2—Continued

| Short Label (1) | Long Label (2) | Units (3) | Description (4) |
|---|---|---|---|
| TBD | (ORIGIN_J) | | origin of photometry |
| TBD | (ORIGIN_H) | | origin of photometry |
| TBD | (ORIGIN_K) | | origin of photometry |
| TBD | (ORIGIN_3p6um) | | origin of photometry |
| TBD | (ORIGIN_4p5um) | | origin of photometry |
| TBD | (ORIGIN_5p8um) | | origin of photometry |
| TBD | (ORIGIN_8p0um) | | origin of photometry |
| **SED Properties** (Povich et al. 2013) | | | |
| TBD | (SED_FLG) | | classification from SED analysis |
| TBD | (SED_AV) | mag | Av from SED analysis |
| TBD | (SED_STAGE) | | YSO stage |
| **X-ray Classification** (Section 3) | | | |
| H1-Prior | (H1_prior) | | class prior probability (position-dependent) |
| H2-Prior | (H2_prior) | | class prior probability (position-dependent) |
| H3-Prior | (H3_prior) | | class prior probability (position-dependent) |
| H4-Prior | (H4_prior) | | class prior probability (position-dependent) |
| H1-Post | (H1_posterior) | | class posterior probability |
| H2-Post | (H2_posterior) | | class posterior probability |
| H3-Post | (H3_posterior) | | class posterior probability |
| H4-Post | (H4_posterior) | | class posterior probability |
| TBD | (H2_dominant_factor) | | dominant classification term[i] |
| Assign | (xray_class_code) | | classification (0=unclassified, 1=H1, 2=H2, 3=H3, 4=H4) |

Note. — Col. (1): Short column label chosen by ApJ and used in the electronic edition of this table. Col. (2): Long column label previously published by the CCCP (Broos et al. 2011a,b) and produced by the *ACIS Extract* (AE) software package (Broos et al. 2010, 2012). The AE software and User's Guide are available at http://www.astro.psu.edu/xray/acis/acis_analysis.html.

The suffixes "_t", "_s", and "_h" on names of X-ray photometric quantities designate the *total* (0.5–8 keV), *soft* (0.5–2 keV), and *hard* (2–8 keV) energy bands.

[a] X-ray source labels identify a *Chandra* pointing; they do not convey membership in astrophysical clusters.

[b] In statistical hypothesis testing, the p-value is the probability of obtaining a test statistic at least as extreme as the one that was actually observed when the null hypothesis is true.

[c] See Broos et al. (2010, Section 7.6) for a description of the variability metrics, and caveats regarding possible spurious indications of variability using the ProbKS_merge metric.

[d] Due to dithering over inactive portions of the focal plane, a *Chandra* source is often not observed during some fraction of the nominal exposure time. (See http://cxc.harvard.edu/ciao/why/dither.html.) The reported quantity is FRACEXPO produced by the *CIAO* tool *mkarf*.

[e] Some background events arising from an effect known as "afterglow" (http://cxc.harvard.edu/ciao/why/afterglow.html) may contaminate source extractions, despite careful procedures to identify and remove them during data preparation (Broos et al. 2010, Section 3). After extraction, we attempt to identify afterglow events using the tool ae_afterglow_report, and report the fraction of extracted events attributed to afterglow; see the *ACIS Extract* manual (http://www.astro.psu.edu/xray/acis/acis_analysis.html).

[f] Confidence intervals (68%) for NetCounts quantities are estimated by the *CIAO* tool *aprates* (http://asc.harvard.edu/ciao/ahelp/aprates.html).

[g] *MedianEnergy* is the *ACIS Extract* quantity ENERG_PCT50_OBSERVED, the median energy of extracted events, corrected for background (Broos et al. 2010, Section 7.3).

[h] *PhotonFlux* = (NetCounts / MeanEffectiveArea / ExposureTimeNominal) (Broos et al. 2010, Section 7.4)

[i] *H2_dominant_factor* reports the classifier term that exerts the most influence on the H2 posterior probability (1 = prior, 2 = MedianEnergy, 3 = $J$ magnitude, 4 = X-ray variability, 5 = spectral type, 6 = 4.5 $\mu$m magnitude, 7 = infrared SED model).

## 2.1. Orion Nebula, Carina Nebula, and W40

Since members of our team have previously published X-ray catalogs for the Orion Nebula (Getman et al. 2005a), Carina Nebula (Broos et al. 2011a), and W 40 (Kuhn et al. 2010) using procedures and software that are very similar to those in MYStIX, we have not constructed new versions of those X-ray catalogs. Since those publications also identified NIR counterparts to those X-ray catalogs, we have not repeated that task.

The X-ray detected entries in the MPCM catalog for the Orion Nebula were obtained from a highly-reliable published membership study of X-ray sources (Getman et al. 2005b). Six additional X-ray sources not previously recognized as members and 110 IR excess sources not detected by *Chandra* were identified by Megeath et al. (2012). NIR counterparts and photometry were reported by Getman et al. (2005b, Section 10) using NIR catalogs from several facilities, including the VLT-ISAAC camera and 2MASS. MIR photometry was obtained from Megeath et al. (2012).

The X-ray detected entries in the MPCM catalog for the Carina Nebula were obtained from the published CCCP study of X-ray members (Broos et al. 2011b), which was the prototype for the X-ray classification procedure described in Section 3. Additional members not detected by *Chandra* were obtained from the IR excess sources identified by Povich et al. (2011). NIR counterparts were identified by Broos et al. (2011b) and NIR photometry (Preibisch et al. 2011) was obtained from the VLT HAWK-I camera.[3] Since the CCCP's deep NIR data cover only a portion of the X-ray field of view (Townsley et al. 2011, Figure 5), we have cropped this MPCM catalog to that NIR field. MIR photometry was obtained from the Vela-Carina Survey (*Spitzer* Proposal ID 40791, PI S. Majewski). OB stars were obtained from the CCCP catalog of massive stars (Gagné et al. 2011).

The X-ray detected entries in the MPCM catalog for W 40 were obtained from a published membership study of X-ray sources (Kuhn et al. 2010). Additional members not detected by *Chandra* were identified by the MYStIX SED modeling procedures. NIR and MIR counterparts and photometry were identified by the MYStIX procedures.

## 2.2. SIMBAD Notes

The MPCM catalog is accompanied by over four thousand footnotes that summarize published information on the 31,000 MSFR members in Table 1. This information is qualitatively different from that used in the MYStIX classification. The most common footnotes are derived from visual-band spectroscopy (spectral type, H$\alpha$ emission), visual-band variability and multiplicity, radio emission (continuum, maser), and submillimeter emission.

Footnote information was obtained from searches of the SIMBAD database (Wenger et al. 2000) within $2^{tt}$ radius around the MYStIX star location, and from the massive star tabulations of Skiff (2009). *It is important to recognize that some of these footnote associations will be incorrect.* No catalog matching algorithm or scientific judgment was applied; in crowded fields, possible multiple counterparts are not resolved and the cited object may not be physically related to the MPCM star. Associations with extended structures, such as dust continuum cloud cores and pillars, will also be incomplete. Multiple designations for the same star are often omitted. Associations are omitted when only X-ray or infrared photometric information is available. Except for a visual magnitude, the footnotes do not provide photometric infor-mation. X-ray/NIR associations suggested by the footnotes are not expected to be fully consistent with the X-ray/NIR associations declared in the MPCM catalog itself.

---

[3] HAWK-I observations were obtained on the ESO 8-meter Very Large Telescope (VLT) at Paranal Observatory, Chile, under ESO programme 60.A-9284(K).

Although the footnote associations are not always reliable, they are useful in several respects. They permit rapid association between MPCM stars and stars that have been scientifically studied as MSFR members. Some MYStIX regions have hundreds of SIMBAD listings, indicating extensive past study of the stellar population, while others have only a handful of SIMBAD associations. Cursory examination of the footnotes gives a sense of the types of previously known stars in the region. Some MSFRs have a considerable population of bright OB stars, others have hundreds of faint H$\alpha$ pre-main sequence stars, a few have massive protostars with masers and compact H II regions, and others have virtually no previous measurements. The footnotes also give guidance to anyone who wishes to construct even larger stellar samples by combining MYStIX sources with the non-MYStIX sources of other surveys.

### 2.3. Spatial Distribution

Figure 1 gives a snapshot of the spatial structure of the MYStIX fields by plotting MPCM stars on *Spitzer* IRAC 8 $\mu$m maps. The symbol colors show the origin of each star—yellow for X-ray detection and red for infrared excess; some stars have both characteristics.

The apparent spatial distributions of disk-bearing (red) and diskless (yellow without concentric red) stars often trace the same clusters. Many of the differences between these distributions are likely to be observational effects. Reduced IRAC point-source sensitivity from IR nebulosity and crowding can produce an apparent decrease in the disk-bearing to diskless ratio, as in the main clusters of RCW 38, NGC 6357, M 17, and NGC 3576. At the edges of the *Chandra* pointings, reduced X-ray point-source sensitivity can produce an apparent increase in the disk-bearing to diskless ratio, e.g., in NGC 6334, NGC 6357, and Trifid. However, clusters in NGC 2264, Rosette, and Eagle exhibit variations in apparent disk fraction that is not easily explained by observational biases.

A variety of cluster structures is apparent. In NGC 2264, DR 21, and NGC 6334 the stellar population is dominated by multiple young clusters embedded in clouds with sinuous structures. In Flame and RCW 36 our field of view contains only single young clusters that are in fact embedded in parts of much larger sinuous clouds. Rosette and Eagle have multiple clusters embedded in clumpy molecular structures next to rich older clusters that have been freed from their parental clouds.

These findings are consistent with a well-accepted model of star formation in turbulent and filamentary giant molecular clouds, in which portions of the cloud, at different times and different locations, exhibit the conditions for gravitational collapse to form a star cluster. The first rich clusters that have OB stars with powerful ultraviolet radiation and winds will evacuate an H II region. Older clusters thus often appear within interstellar bubbles and may be less absorbed than younger clusters forming deep in other portions of the cloud. It is unclear whether the recent star formation is triggered by the expanding H II regions or whether it is occurring spontaneously in dense cloud filaments. Both processes are likely to be present.

Kuhn et al. (2013c) (in preparation) will model star density concentrations as isothermal ellipsoids, allowing stars to be associated with clusters and subclusters in an objective manner. Other MYStIX studies will follow, discussing of cluster ages, relationships between cluster properties, small-scale clustering, and mass segregation.

– 11 –

Flame Nebula, RCW 36, NGC 2264

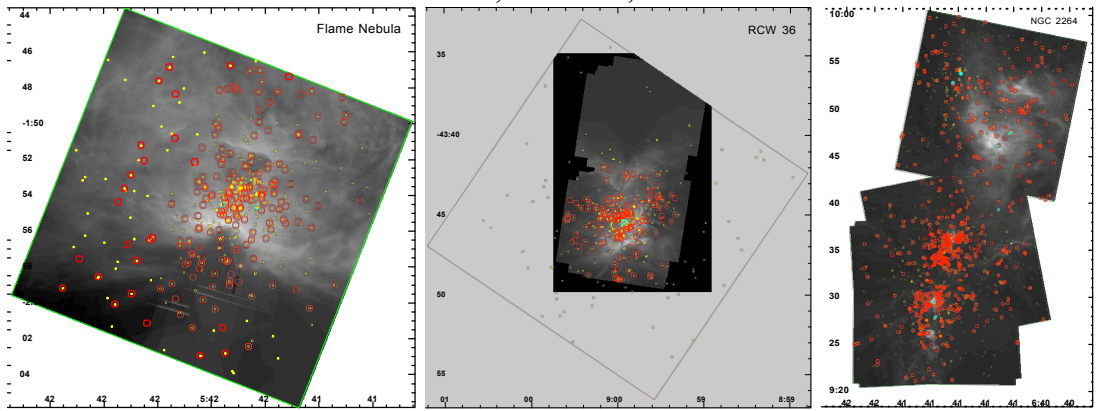

Rosette Nebula, Lagoon Nebula

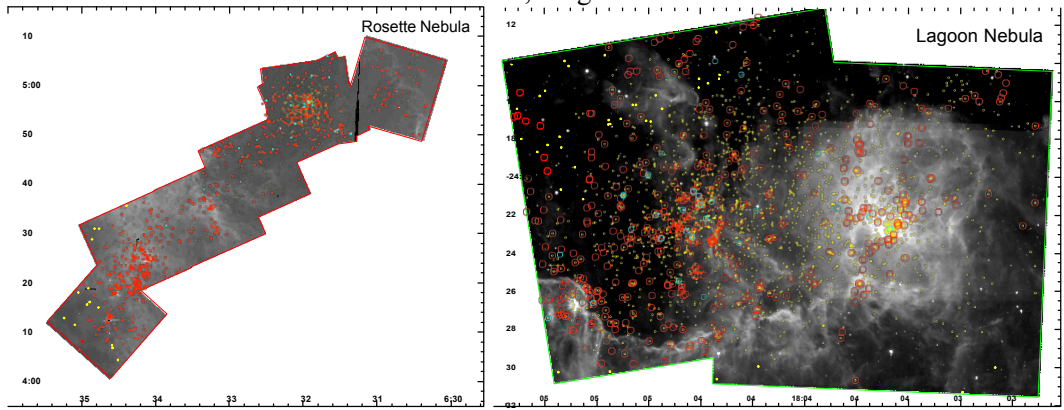

NGC 2362, DR 21, RCW 38

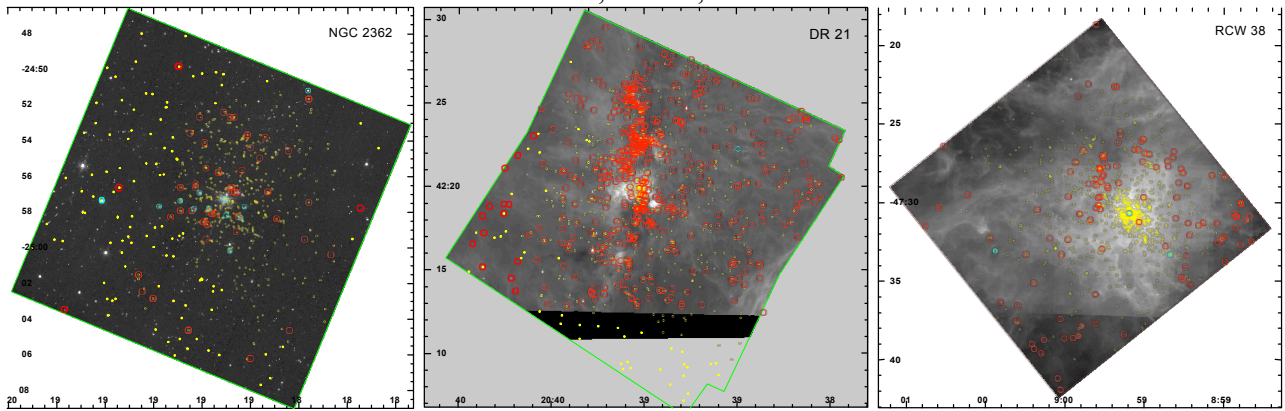

NGC 6334

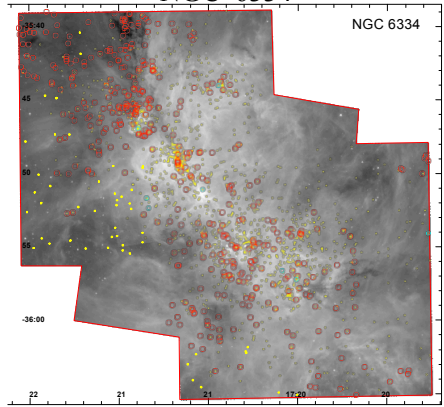

Fig. 1.—: X-ray sources (yellow), infrared excess sources(red), and massive stars (cyan) from the MPCM catalog, shown on an 8 $\mu$m *Spitzer* image. Each Chandra exposure subtends $17^t \times 17^t$; most fields are mosaics of several pointings.

## 3. Classification of X-ray Sources

The MYStIX *Chandra* observations have sufficient angular resolution and sensitivity to detect hundreds to thousands of young stars in each region; their X-ray emission arises mainly from magnetic reconnection and from massive star winds. Older field stars (foreground and background) and extragalactic objects are also detected. These object classes produce distinct observed X-ray spectra (due to different emission mechanisms and absorption column densities) and distinct patterns of X-ray variability. In principle, analysis of X-ray spectra and light curves could distinguish MSFR members from contaminants. However, *Chandra*'s exquisite angular resolution and low background produce X-ray catalogs in which the vast majority of sources have far too few counts to achieve this goal.

An infrared counterpart can significantly clarify the classification of an X-ray source. For example, extragalactic X-ray sources in MYStIX catalogs are rarely brighter than 20th magnitude in the $J$ band (Alexander et al. 2001) or 13th magnitude in the 4.5 $\mu$m band (Harvey et al. 2007). Since disk-bearing young stars, field stars, and galaxies have distinct infrared SEDs, analysis of multi-band infrared photometry can provide classification evidence. When an X-ray source is reliably identified as a massive star, one can reasonably conclude that it is a member of the MSFR, since massive field stars are very rare. The position of a source in the MYStIX field of view suggests its class—sources lying close to cluster centers are clearly more likely to be members than stars in more outlying regions.

Several measured source properties exhibit member and contaminant distributions that overlap significantly, but have distinct shapes. Traditional classification decision trees that involve thresholding measurements are inappropriate for these data, due to the overlapping distributions. However, the distinct shapes of the member and contaminant distributions indicate that these measurements do carry information relevant to classification. Thus, we wish to interpret measurements as "weighted evidence" for source class, rather than as inputs to a classification decision tree.

Furthermore, in order to infer a classification for each of our X-ray sources or to conclude that the classification is undetermined, we require a clearly-defined framework for combining whatever observations are available. Note that the observations may be in conflict for a particular source; for example, the X-ray spectrum may be most consistent with a young star in the MSFR but the $J$ magnitude may be most consistent with a background star. We wish to resolve such conflicts via a consistent and principled procedure.

### 3.1. A Classification Framework

We adopt a "Naive Bayes Classifier" (Duda et al. 2002) that is closely related to the X-ray source classifier used in the CCCP (Broos et al. 2011b). The Naive Bayes approach has advantages for our problem. First, this type of classifier provides, for each source, real-valued class probabilities, not just class decisions. Thus, the rule we chose for deciding when a source cannot be reliably classified (Section 3.3) can be easily replaced by more conservative or more liberal policies in subsequent studies, to strike a different balance between classification accuracy and completeness. Second, this type of classifier is applicable when the source properties considered by the classifier are very different (e.g., $J$ magnitude, X-ray variability, and the categorical presence of an IR-excess). Some other classification methods require construction of a "distance metric" between objects in a multi-dimensional measurement space, which requires the specification of arbitrary scaling relationships among variables with incompatible units.

We briefly present the MYStIX X-ray classification model here; Broos et al. (2011b) describe the CCCP classifier in more detail. First, we define a set of mutually exclusive classification hypotheses, denoted as {**H1,H2,H3,H4**}, that represent the four types of objects that *Chandra* detects in MYStIX observations: young stars in the MSFR, and three populations of contaminants.

**H1:** source is a foreground Galactic field star

**H2:** source is a young star in the MSFR

**H3:** source is a background Galactic field star

**H4:** source is an extragalactic object

Second, for each contaminant population (H1, H3, H4) we create a map of the expected density of detected X-ray sources across the *Chandra* field of view, by applying adaptive kernel smoothing to a two-dimensional histogram of the positions of simulated contaminants (Section 3.2). We also create a density map for the observed catalog of X-ray sources, $\rho_{\rm obs}(\mathbf{r})$, via adaptive smoothing. These four density maps—$\rho_{H1}(\mathbf{r})$, $\rho_{H3}(\mathbf{r})$, $\rho_{H4}(\mathbf{r})$, $\rho_{\rm obs}(\mathbf{r})$—are functions of celestial position, $\mathbf{r}$, and have units of detected X-ray sources per unit area on the sky. The simple arithmetic in Equation 1 transforms those surface density maps into unitless maps representing the *fraction* of observed X-ray sources expected to belong to each class, based only on source location.

$$\begin{aligned}
{\rm prior}_{H1}(\mathbf{r}) &= \rho_{H1}(\mathbf{r}) / \rho_{\rm obs}(\mathbf{r}) \\
{\rm prior}_{H2}(\mathbf{r}) &= (\rho_{\rm obs}(\mathbf{r}) - \rho_{H1}(\mathbf{r}) - \rho_{H3}(\mathbf{r}) - \rho_{H4}(\mathbf{r})) / \rho_{\rm obs}(\mathbf{r}) = 1 - \frac{\rho_{H1}(\mathbf{r}) + \rho_{H3}(\mathbf{r}) + \rho_{H4}(\mathbf{r})}{\rho_{\rm obs}(\mathbf{r})} \\
{\rm prior}_{H3}(\mathbf{r}) &= \rho_{H3}(\mathbf{r}) / \rho_{\rm obs}(\mathbf{r}) \\
{\rm prior}_{H4}(\mathbf{r}) &= \rho_{H4}(\mathbf{r}) / \rho_{\rm obs}(\mathbf{r}).
\end{aligned} \quad (1)$$

When evaluated at the position of a source in our X-ray catalog, these four class fractions sum to unity (${\rm prior}_{H1} + {\rm prior}_{H2} + {\rm prior}_{H3} + {\rm prior}_{H4} = 1$) and represent a set of "prior probabilities" for the source's class.

Our use of the word "prior" here is an intentional reference to its meaning in Bayesian inference, namely the probability of a hypothesis prior to consideration of the measurements at hand. In our formulation presented here, the observed position of a source is an input to the calculation of the class prior probabilities for that source; the position is not interpreted as a "measurement". However, Broos et al. (2011b) provide an appendix that defines an equivalent formulation in which a single set of (position-independent) prior class probabilities apply to the entire catalog, and the position of an individual source is interpreted as a measurement.

Third, we define four probability density functions (PDFs), $p(D_1, D_2, \ldots D_N \,|\, {\rm class} = H)$, that express for an individual source the probability of obtaining specific measurements for a set of $N$ source properties $(D_1, D_2, \ldots D_N)$, conditioned on the source belonging to a specific class ($H \in \{H1, H2, H3, H4\}$). These measured source properties may be a mix of continuous (e.g., Section 3.2.1, 3.2.2, 3.2.5) and discrete (e.g., Section 3.2.3, 3.2.4, 3.2.6) quantities. These PDFs encode our understanding of what combinations of data are produced by sources from each of our four parent populations. In principle, they encapsulate our understanding of the physics of the X-ray and infrared emission of these four classes of objects, plus observational effects such as absorption, instrument response, and survey sensitivities. Mathematically, these are $N$-dimensional joint PDFs that represent the many physical correlations that exist among the observable source properties.

The Naive Bayes classifier makes the common and critical simplification that these joint PDFs can be approximated by the product of one-dimensional PDFs. More formally, it assumes that the observed properties of a source are statistically independent:

$$p(D_1, D_2, \ldots D_N \,|\, {\rm class} = H) = p(D_1 \,|\, {\rm class} = H) \, p(D_2 \,|\, {\rm class} = H) \ldots p(D_N \,|\, {\rm class} = H)$$

$$= \prod_{i=1}^{N} p(D_i \,|\, \text{class} = H). \qquad (2)$$

Each term on the right-hand side of Equation 2 is the expected distribution of a single source property, say $J$-band magnitude, for sources in the class $H$. These $4N$ one-dimensional PDFs are estimated in Section 3.2.

For a particular X-ray source, the variables $D_1, D_2, \ldots D_N$ in Equation 2 have specific values obtained from observations. With $D_1, D_2, \ldots D_N$ fixed, Equation 2 can be viewed as a function of the discrete variable $H$ (appearing in the condition "class $= H$"), representing the source class hypothesis (H1, H2, H3, or H4). In the field of statistics, such a function is formally called a "likelihood function." We adopt that terminology here, sometimes using the shorthand "class likelihood." In summary, each X-ray source produces four class likelihood values by evaluating Equation 2 for $H = H1 \ldots H4$. We will discuss these likelihood functions in more detail in Section 3.2.

The assumption of independence is not strictly correct in the X-ray source classification problem. For example, a harder X-ray spectrum will be somewhat correlated with fainter $J$ magnitude, as both are products of heavier obscuration. Also, for MSFR members, the detection of rapid X-ray variability will be correlated with X-ray flux, which itself is linked to pre-main sequence stellar mass and $J$ magnitude (Telleschi et al. 2007). However, Naive Bayes is often surprisingly good even when the independence assumption is violated (Hand & Yu 2001).

As with many multivariate problems, we must treat the case of missing data. When an estimate of a property is not available for a specific X-ray source, we choose to omit that term from Equation 2. The missing source property thus plays no role in the classification decision.

Finally, Bayes' Theorem provides a coherent and conceptually simple method for combining prior probabilities for hypotheses and likelihood functions for those hypotheses to produce *posterior probabilities* for those hypotheses, conditioned on the data we have observed:

$$\text{posterior} \propto \text{likelihood} \times \text{prior}. \qquad (3)$$

$$p(H \,|\, D_1, D_2, \ldots D_N) \propto p(D_1, D_2, \ldots D_N \,|\, H) \, p(H)$$

For a source at location $\mathbf{r}$ with observed source properties $D_1, D_2, \ldots D_N$, the posterior probabilities for the four possible values of $H$ can be written as

$$\text{Prob}(\text{class} = H1 \,|\, \mathbf{r}, D_1, D_2, \ldots D_N) = k \prod_{i=1}^{N} p(D_i \,|\, \text{class} = H1) \, \text{prior}_{H1}(\mathbf{r})$$

$$\text{Prob}(\text{class} = H2 \,|\, \mathbf{r}, D_1, D_2, \ldots D_N) = k \prod_{i=1}^{N} p(D_i \,|\, \text{class} = H2) \, \text{prior}_{H2}(\mathbf{r})$$

$$\text{Prob}(\text{class} = H3 \,|\, \mathbf{r}, D_1, D_2, \ldots D_N) = k \prod_{i=1}^{N} p(D_i \,|\, \text{class} = H3) \, \text{prior}_{H3}(\mathbf{r})$$

$$\text{Prob}(\text{class} = H4 \,|\, \mathbf{r}, D_1, D_2, \ldots D_N) = k \prod_{i=1}^{N} p(D_i \,|\, \text{class} = H4) \, \text{prior}_{H4}(\mathbf{r}) \qquad (4)$$

The common constant of proportionality, k, is easily found by requiring that the four posterior probabilities sum to unity.

## 3.2. Estimating Likelihood Terms in the Classifier

The inputs to the MYStIX posterior class probability calculation for a specific X-ray source (right-hand sides of Equation 4) require up to seven numbers for each class: a class prior probability and class likelihood values for up to six observed source features that we have chosen for classification purposes. Table 3 lists those seven inputs (Col. 1), the conceptual role they play in Bayes Theorem (Col. 2), and the strategy we used to estimate their values for the four classes (Cols. 3–6).

Two of our six observed source features are continuous quantities: median X-ray energy (Section 3.2.1) and $J$ magnitude (Section 3.2.2). Two other continuous quantities—a statistic related to X-ray variability (Section 3.2.3) and 4.5 $\mu$m magnitude (Section 3.2.5)—are quantized during evaluation of their likelihood functions. The remaining two source features—spectral type (Section 3.2.4) and SED classification (Section 3.2.6)—are intrinsically discrete quantities.

Astrophysical simulations of objects in the three contaminant classes—tailored to each region's *Chandra* observation parameters, Galactic sight-line, distance, foreground absorption, and background absorption—play a vital role (Getman et al. 2011). These simulations provide source density maps for the contaminant classes (Appendix A), which are used in the calculation of the class priors (Equation 1). They also provide two sets of likelihood functions for the contaminant classes (H1, H3, H4) in the form of PDFs of two important source measurements: median X-ray energy (Section 3.2.1) and $J$ magnitude (Section 3.2.2).

We judge that obtaining the median energy and $J$ magnitude likelihood functions for the member class (H2) from simulations is not feasible because too many critical astrophysical assumptions would have to be made, including the spatial, mass, age, and absorption distributions of detectable members. Instead, we choose to use empirical PDFs obtained from a subsample of X-ray sources that are almost certainly members. Such a set of objects with a known or presumed classification is commonly referred to as a "training set." Since the training set is constructed using a simplified version of the classifier itself, we postpone further discussion of the training set construction until Section 3.2.7, after the likelihood terms in the classifier have been defined.

Table 3. Components of the Classification Model

|  | Observation | Role | H1 foreground | H2 member | H3 background | H4 extragalactic |
|---|---|---|---|---|---|---|
|  | (1) | (2) | (3) | (4) | (5) | (6) |
|  | source density | prior (Eqn. 1) | simulation | observation | simulation | simulation |
| $D_1 =$ | median X-ray energy | likelihood | simulation | training set | simulation | simulation |
| $D_2 =$ | $J$ magnitude | likelihood | simulation | training set | simulation | simulation |
| $D_3 =$ | X-ray variability | likelihood | theory | training set | theory | theory |
| $D_4 =$ | visual spectroscopy | class veto | ⋯ | judgment | ⋯ | ⋯ |
| $D_5 =$ | 4.5 $\mu$m magnitude | class veto | ⋯ | ⋯ | ⋯ | Harvey et al. (2007) |
| $D_6 =$ | infrared SED model | likelihood | judgment | judgment | judgment | judgment |

– 17 –### 3.2.1. Likelihood for median X-ray energy

Foreground stars will generally have lower line-of-sight absorption than MSFR members, whereas background stars and extragalactic sources may have higher absorption. The nature of the emitting plasma is also expected to differ among these populations. For example, the intrinsic X-ray spectrum of older field stars will be cooler (similar to our Sun's "coronal" emission) than that of young members of a star-forming region, which is dominated by magnetic reconnection flare emission (e.g. Güdel & Nazé 2009). In combination, these two factors produce distinct shapes for the typical apparent X-ray spectra from these populations, which can be characterized by distinct distributions for the median energy of detected X-ray photons (Getman et al. 2010). The median statistic is chosen because it is robust against outliers and is available for sources with few detected photons.

Figure 2 shows an example of estimated median X-ray energy PDFs, conditioned on each of the four source classes. In the region shown, the Trifid Nebula, the PDFs for each class are particularly well separated by absorption along the line of sight. Since members (red) are distant (d=2.7 kpc), they appear significantly harder than foreground stars (black). Beyond the MSFR, the sightline ($l, b = 7.0°, -0.3°$) includes many obscured background stars (green) towards the Galactic Center. Similarly, spectra from extragalactic sources (blue) are hardened further by their sightline through much of the Galactic disk. These PDFs were constructed specifically for the X-ray observation of the Trifid Nebula by applying adaptive kernel smoothing to sample histograms of median X-ray event energies obtained from contaminant simulations (Section 3.2) and from the member training set (Section 3.2.7). If a source has a reliable median energy estimate ($\geq 4$ net events detected), then the four PDF values at that energy form the class likelihoods for this term in Equation 4.

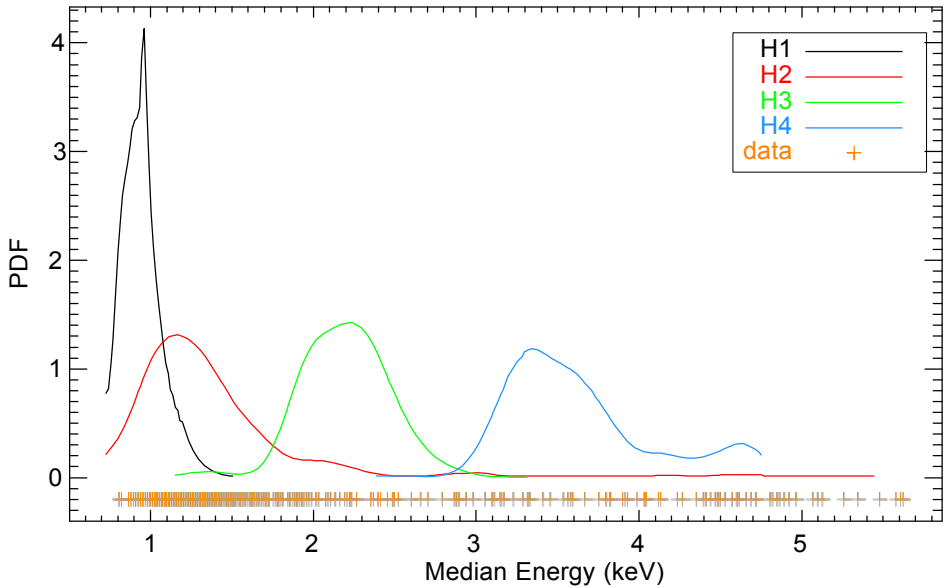

Fig. 2.—: Trifid class likelihood functions for median X-ray energy: H1=black, H2=red, H3=green, H4=blue. Measured median X-ray energies for X-ray sources (orange +) are marked below the functions. Corresponding figures for each MSFR listed in Table 8 are available in the electronic edition of this article.

*3.2.2. Likelihood for J magnitude*

Foreground stars, MSFR members, and background stars in an X-ray sample often exhibit somewhat different distributions in $J$ magnitude, due to distance and absorption. More importantly, extragalactic sources typically have distinctly fainter $J$ magnitude than the stars we detect.

Figure 3 shows the conditional $J$ magnitude PDFs for the MSFR (NGC 6334) in which they are most well separated. Although the distributions expected for foreground stars (black), MSFR members (red), and background stars (green) overlap significantly, at many $J$ magnitudes the ratio between the largest and smallest PDF is large, providing strong classification evidence. These PDFs were constructed specifically for the X-ray observation of NGC 6334 by applying adaptive kernel smoothing to sample histograms of $J$ magnitude obtained from contaminant simulations (Section 3.2) and from the member training set (Section 3.2.7).

Each observed X-ray source with an identified NIR counterpart (Naylor et al. 2013) produces four class likelihoods for this term in Equation 4. The absence of a $J$-band counterpart is **not** interpreted as evidence for any source class (e.g., the extragalactic class); for such cases we drop the $J$-band term from Equation 4. We feel it would be incorrect to interpret a missing $J$ value as an upper limit because the sensitivity of the NIR surveys can vary spatially due to emission nebulosity, cloud obscuration, or proximity to a bright star.

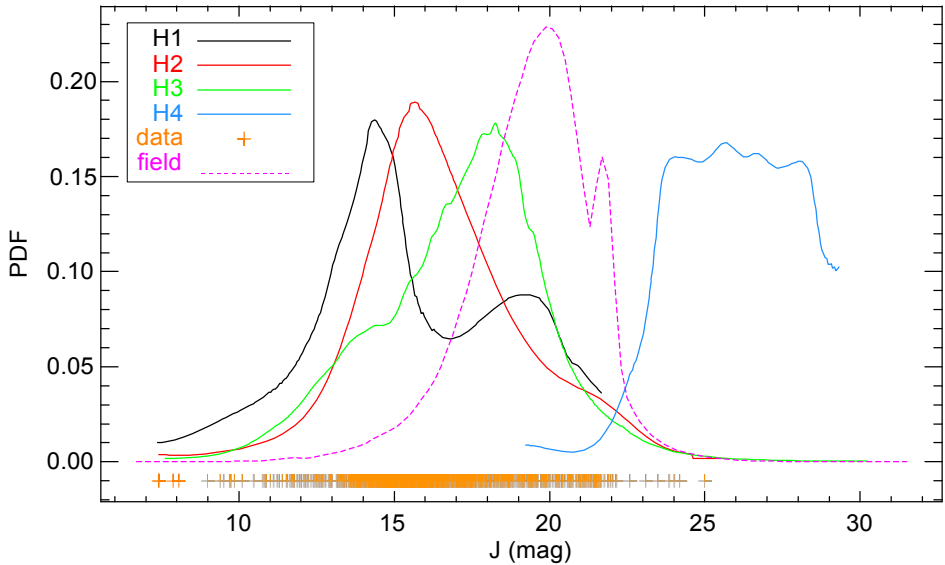

Fig. 3.—: NGC 6334 class likelihood functions for J magnitude: H1=black, H2=red, H3=green, H4=blue. Measured J magnitudes for X-ray sources (orange +) are marked below the functions. J magnitudes for all objects in the field are shown (purple, dotted) to indicate field-averaged NIR completeness. Corresponding figures for each MSFR listed in Table 8 are available in the electronic edition of this article.

### 3.2.3. Likelihood for X-ray variability

High amplitude, rapid X-ray flares are frequent in young stars but are less common in older stars (Wolk et al. 2005). While nearly all extragalactic sources exhibit variability on timescales of days to months in the *Chandra* band, only a small fraction (< 15%) have detectable variations within 1 day (Paolillo et al. 2004; Shemmer et al. 2005). Thus, X-ray variability is a powerful classification input.

In the X-ray catalog, variability is quantified by a p-value[4] for the no-variability hypothesis, estimated via the Kolmogorov-Smirnov (K-S) statistic (ProbKS_single in Broos et al. 2011a). In the classifier, this p-value is discretized into a variability grade with three values, defined in Table 4. We assume the null hypothesis is true for the contaminant populations (H1,H3,H4), i.e. that those sources are constant on the timescale of single *Chandra* observations. Thus, the expected distribution of the variability grade in those classes is, by definition, obtained from the p-values used to define the grade, as shown in Table 4. For example, the "definitely variable" grade is assigned when a p-value < 0.005 is found for the K-S statistic; for sources with constant flux this will occur by chance with a probability of 0.005, and thus the likelihoods for the "definitely variable" observation are assigned that value for the H1, H3, and H4 classes. The expected distribution of the variability grade for MSFR members (H2 class) is estimated separately for each region from the corresponding H2 training set (Section 3.2.7); example values for one MYStIX MSFR (M 17) are shown in Table 4.

For all MSFRs, a variability grade of "definitely variable" is interpreted as strong evidence for the H2 class; the grade "possibly variable" is interpreted as moderate evidence for the H2 class. Because the variability PDF for each class is normalized (as required by the definition of a PDF), the grade "no evidence" is logically interpreted as evidence *against* the H2 class (Broos et al. 2011b).

---

[4] In statistical hypothesis testing, the p-value is the probability of obtaining a test statistic (such as the Kolmogorov-Smirnov statistic) at least as extreme as the one that was actually observed when the null hypothesis is true.

Table 4. Likelihood functions for variability grades in M 17

| Grade | | Definition | Prob($D_3|H$), from Eqn. 4 | |
|---|---|---|---|---|
| | | | H1,H3,H4 | H2[a] |
| No evidence for variability | ⇔ | 0.05 < ProbKS_single | 0.950 = 1.00 − 0.05 | 0.57 |
| Possibly variable | ⇔ | 0.005 < ProbKS_single < 0.05 | 0.045 = 0.05 − 0.005 | 0.21 |
| Definitely variable | ⇔ | ProbKS_single < 0.005 | 0.005 | 0.22 |

[a]The variability PDF for the H2 class is MSFR-dependent.

### 3.2.4. Likelihood for visual spectroscopy

Although massive stars (spectral types B3 or earlier, identified by visual spectroscopy) are explicitly added to the MPCM list, whether detected by *Chandra* or not (Section 1), we also choose to use spectral type in the X-ray classification model so that the H2 training sets (Section 3.2.7) are guaranteed to include these stars. Operationally, an early spectral type vetoes the H1, H3, and H4 classes by setting their likelihoods to zero; all of the posterior probability is forced into the H2 class. This likelihood term is omitted when the spectral type is not known to be massive. These policies are summarized in the $D_4$ section of Table 5.

### 3.2.5. Likelihood for 4.5 µm magnitude

Harvey et al. (2007) used the observed distribution of fluxes in one of the *Spitzer* Wide-Area Infrared Extragalactic survey (SWIRE) fields to place an upper limit on the 4.5 µm flux produced by extragalactic point sources detected by *Spitzer*. As in the CCCP classifier, we interpret a bright 4.5 µm magnitude ($[4.5] < 13$ mag) as a veto of the H4 class, implemented by setting this term's H4 likelihood to zero. This criterion is conservative, in that extragalactic sources observed through the obscuration of the MYStIX sightlines are expected to be even fainter than in the SWIRE field. Since we have no models for the full distribution of [4.5] within the four classes, no preference among H1, H2, and H3 is expressed when H4 is vetoed (i.e. H1, H2, and H3 are assigned equal likelihoods), and this likelihood term is omitted when 4.5 µm magnitude is faint ($[4.5] > 13$ mag). These policies are summarized in the $D_5$ section of Table 5.

### 3.2.6. Likelihood for infrared SED Model

Recall from Section 1 that one of the three ways in which we identify MPCM sources is SED modeling. An IR source *not detected by Chandra* can enter the MPCM list when an IR excess strongly indicating circumstellar dust in a disk or infalling envelope is found (Povich et al. 2013). Those authors have a separate IR SED analysis procedure that is attempted on all X-ray sources; it takes advantage of the fact that X-ray detection is very unlikely among the dominant contaminating population in an infrared-selected SED analysis. That analysis produces one of several inferences about the astrophysical object that generated the SED, listed below. We used our professional judgment to decide how those SED inferences should be interpreted as classification evidence, as shown in the $D_6$ section of Table 5.

- We interpret the inference of "likely YSO" as strong but not certain evidence of MSFR membership, represented by a 30-to-1 likelihood ratio between H2 and each of the other classes. The equal likelihoods for the H1, H3, and H4 classes represent our inability to quantify the small fraction of objects in those classes with SEDs that look like YSOs.

- We judge that the "stellar photosphere" and "marginal IR excess" inferences constitute certain evidence against extragalactic sources, represented by a zero likelihood for the H4 class, and we assert that these inferences should play no role in choosing among the other classes, represented by equal likelihoods for the H1, H2, and H3 classes.

- We judge that the "candidate galaxy/PAH" and "candidate AGN" inferences constitute certain evidence against foreground and background stars, represented by zero likelihoods for the H1 and H3 classes, and we judge that these inferences favor the extragalactic class over the MSFR member class, represented by the 2-to-1 likelihood ratio between H4 and H2.

Table 5. Likelihood functions for source properties $D_4$, $D_5$, $D_6$

| Observed Property | Prob($D_i \mid H$) | | | |
|---|---|---|---|---|
| | H1 | H2 | H3 | H4 |
| $D_4$: Spectral Type | | | | |
| B4 or later | ⋯ | ⋯ | ⋯ | ⋯ |
| B3 or earlier | 0 | 1 | 0 | 0 |
| $D_5$: [4.5 $\mu$m] | | | | |
| < 13 mag | 1 | 1 | 1 | 0 |
| > 13 mag | ⋯ | ⋯ | ⋯ | ⋯ |
| $D_6$: SED Model (Povich et al. 2013) | | | | |
| likely YSO | 1 | 30 | 1 | 1 |
| stellar photosphere | 1 | 1 | 1 | 0 |
| marginal IR excess | 1 | 1 | 1 | 0 |
| candidate galaxy/PAH | 0 | 1 | 0 | 2 |
| candidate AGN | 0 | 1 | 0 | 2 |
| No well-fit models | ⋯ | ⋯ | ⋯ | ⋯ |
| No fit attempted | ⋯ | ⋯ | ⋯ | ⋯ |

### 3.2.7. Construction of an H2 Training Set

As shown in Table 3, the median X-ray energy, $J$ magnitude, and X-ray variability likelihood terms cannot be enabled in the classifier until an H2 training set has been constructed and the H2 likelihood functions for those three terms have been constructed. Members of the H2 training set are chosen in three steps. First, we run the classifier with the median energy, $J$ magnitude, and X-ray variability terms disabled to compute preliminary H2 posterior probabilities. Second, we exclude from the training set sources whose H2 posterior probability does not exceed a conservative threshold (0.86), chosen using NGC 2264, where extensive prior knowledge of the young stellar distribution is available (Feigelson et al. 2013).

Among the sources that meet the H2 posterior requirement, a small fraction have little direct evidence that the source is a member. Some have no likelihood terms at all, just the position-dependent prior. To be conservative, our third step requires that training set sources must meet at least one of the following criteria, which represent significant evidence of membership.

1. The source is located within a very dense cluster (prior$_{H2}$ > 0.95).

2. The source is a massive star.

3. The X-ray variability grade is "definitely variable."

4. A 4.5 $\mu$m magnitude is available, and it strongly favors the H2 class over the H4 class, which is our dominant contaminant, i.e., ([4.5] < 13 mag).

5. A $J$ magnitude is available, and it strongly favors the H2 class over the H4 class, which is our dominant contaminant, i.e., ($J$ < 20 mag).

6. A well-fit model was obtained from the IR SED fitting process.

Table 6 (Col. 2) reports the fraction of X-ray sources accepted into the training set for each MSFR. After H2 likelihood functions for median X-ray energy, $J$ magnitude, and X-ray variability are constructed from the training set, the classifier is run a second time (with all six likelihood terms enabled) and we make final class assignments, as described in Section 3.3. This final classification declares a larger fraction of sources to be MSFR members (Col. 3) than were in the training set. Note that membership in the training set is not considered in the final classification run; up to two percent of training set members are not declared to be MSFR members (Col. 5). Col. 4 in Table 6 will be discussed in Section 3.3.

Table 6. Membership Fraction in X-ray Catalog

| MSFR | MSFR members declared | | overturned | TS[a] not confirmed |
|---|---|---|---|---|
| | $\frac{|TS|}{|Xcat|}$[b] | $\frac{|H2|}{|Xcat|}$[c] | $\frac{|H2\text{ overturned}|}{|Xcat|}$[c] | $\frac{|(\text{not}H2)\text{ AND }TS|}{|TS|}$ |
| (1) | (2) | (3) | (4) | (5) |
| Flame Nebula | 67% | 77% | 16% | 0% |
| RCW 36 | 46% | 67% | 22% | 1% |
| NGC 2264 | 56% | 68% | 18% | 0% |
| Rosette Nebula | 47% | 68% | 12% | 1% |
| Lagoon Nebula | 45% | 75% | 12% | 1% |
| NGC 2362 | 34% | 68% | 12% | 0% |
| DR 21 | 46% | 78% | 17% | 2% |
| RCW 38 | 58% | 80% | 11% | 2% |
| NGC 6334 | 69% | 92% | 7% | 1% |
| NGC 6357 | 66% | 83% | 11% | 1% |
| Eagle Nebula | 51% | 73% | 18% | 1% |
| M 17 | 46% | 76% | 5% | 1% |
| W 3 | 46% | 75% | 11% | 2% |
| W 4 | 38% | 64% | 19% | 1% |
| Trifid Nebula | 30% | 66% | 18% | 2% |
| NGC 3576 | 42% | 74% | 20% | 1% |
| NGC 1893 | 47% | 77% | 7% | 2% |

[a]TS refers to the H2 training set (Section 3.2.7).

[b]$|Xcat|$ is the total number of X-ray sources.

[c]Declared and overturned H2 sources are discussed in Section 3.3.

## 3.3. Decision Rule

Probability theory cannot specify how posterior class probabilities should be used for astrophysical analyses; investigators must make that judgment themselves, just as they decide what signal-to-noise ratio is the appropriate threshold in Gaussian detection problems. We choose to adopt a class decision rule that assigns a specific class if the largest posterior probability is more than twice the next-largest posterior probability. When no classification posterior probability stands above the others using this criterion, a source is labeled "unclassified." Although this rule could identify MSFR members that have H2 posteriors as low as 0.40 (i.e. when the H1, H3, and H4 posteriors were all equal to 0.20), in practice 95% to 99.3% of sources declared to be MSFR members have an H2 posterior greater than 0.70.

A small fraction of sources that meet the H2 criterion above have little or no observational evidence, beyond their position-dependent prior, that pertains to classification. To be conservative, we overturn an H2 classification—declare the source to be "unclassified"—if *none* of the following criteria are met:

- Criteria 1, 2, 3, or 6 shown in Section 3.2.7.

- A reliable MedianEnergy is available.

- A $J$ magnitude is available.

Summarizing the criteria above, a source with an overturned H2 classification is not in a cluster core, is not known to be massive, is a weak X-ray detection (MedianEnergy is missing only for sources with less than four net X-ray events), lacks an identified $J$-band counterpart, and has no IR SED analysis. Table 6 (Col. 4) reports the fraction of X-ray sources that are in this "overturned H2" category for all MYStIX regions.

## 4. Results of X-ray Classification

The X-ray classification described in Section 3 was performed on all MYStIX MSFRs except the Orion Nebula, Carina Nebula, and W 40. For these regions we adopted X-ray classifications previously published (Section 2.1).

Table 1 reports tallies of the five possible X-ray classification outcomes: H1 (foreground star), H2 (MSFR member), H3 (background star), H4 (extragalactic), and "unclassified." Figure 4 shows the spatial distribution of all X-ray sources in M 17, color-coded by these classifications. Corresponding figures for each MYStIX region are available in the electronic edition of this article.

The western pointing of M 17 contains the well-studied massive cluster NGC 6618 (Chini & Hoffmeister 2008; Broos et al. 2007). Where the detected source density is very high, virtually every source is classified as a member. Where the source density is moderate in this pointing, a few sources with properties strongly inconsistent with membership are inferred to be foreground (purple), extragalactic (blue), or unclassified (yellow). Most detected sources in this long-exposure (300 ks) pointing have sufficient counts to reliably estimate MedianEnergy,[5] eliminating the so-called "overturned H2" outcome (cyan, Section 3.3). In contrast, the shallow eastern (85 ks) and north-eastern (40 ks) pointings contain many low-count X-ray sources and the "overturned H2" outcome (cyan) is more common. The sparse YSO population in these pointings (Povich et al. 2009) produces a relatively low (∼60 %) member (H2) prior probability, and absorption along the line-of-sight produces well-separated MedianEnergy likelihood functions (M 17 panel in Figure 2). These effects allow significant numbers of foreground (purple) and background (green) stars to be confidently identified (Table 8).

Table 7 defines the columns of an electronic table, available in ASCII format from the electronic edition of this article, that reports prior and posterior class probabilities and the class assignment for every X-ray source (with "0" representing "unclassified"). This table also reports infrared counterpart information (Naylor et al. 2013; King et al. 2013; Kuhn et al. 2013b) for every X-ray source. X-ray properties are available in tables presented by Townsley et al. (2013) and Kuhn et al. (2013a).

---

[5] Sources with very few counts cannot be detected above the high instrumental background arising from this pointing's long exposure time.

Table 7. Counterparts to X-ray Sources and Classification Probabilities

| Column Label (1) | Units (2) | Description (3) |
| --- | --- | --- |

[Column 1 shows the labels used in the FITS table submitted to the ApJ. These labels have not appeared in the literature and will be replaced with short labels that the ApJ will compose.]

**MYStIX coordinates** (Section 2)

| | | |
| --- | --- | --- |
| (MYSTIX_SFR) | ⋯ | MSFR name |
| (Class_Name) | ⋯ | IAU source name; prefix is ___(TBD)??? J |
| (Class_RAdeg) | deg | right ascension (J2000) |
| (Class_DEdeg) | deg | declination (J2000) |
| (Class_Pos_Err) | arcsec | 1-$\sigma$ error circle around (RAdeg,DEdeg) |
| (Class_Pos_Origin) | | origin of position |

**Multi-wavelength Detections**

| | | |
| --- | --- | --- |
| (Xray_Name) | ⋯ | X-ray source name in IAU format |
| (Xray_Label[a]) | ⋯ | X-ray source name used within the MYStIX project |
| (NIR_NAME) | | name in NIR catalog |
| (NIR_LABEL) | | label in NIR catalog |
| (MIR_NAME) | | name in MIR catalog |
| (MIR_LABEL) | | label in MIR catalog |
| (OB_LABEL) | | label in OB catalog |

**IR Counterparts and Photometry**

| | | |
| --- | --- | --- |
| (XN_PROB_CP) | | counterpart probability, X-ray/NIR (Naylor et al. 2013) |
| (XM_PROB_CP) | | counterpart probability, X-ray/MIR (Naylor et al. 2013) |
| (MAG_J) | mag | photometry |
| (ERROR_J) | mag | 1-sigma uncertainty |
| (MAG_H) | mag | photometry |
| (ERROR_H) | mag | 1-sigma uncertainty |
| (MAG_K) | mag | photometry |
| (ERROR_K) | mag | 1-sigma uncertainty |
| (MAG_3p6um) | mag | photometry |
| (ERROR_3p6um) | mag | 1-sigma uncertainty |
| (MAG_4p5um) | mag | photometry |
| (ERROR_4p5um) | mag | 1-sigma uncertainty |
| (MAG_5p8um) | mag | photometry |
| (ERROR_5p8um) | mag | 1-sigma uncertainty |
| (MAG_8p0um) | mag | photometry |
| (ERROR_8p0um) | mag | 1-sigma uncertainty |
| (J_FLAG) | | UKIRT photometry flag (King et al. 2013) |
| (H_FLAG) | | UKIRT photometry flag (King et al. 2013) |
| (K_FLAG) | | UKIRT photometry flag (King et al. 2013) |
| (CC_FLG) | | 2MASS photometry flag |
| (PH_QUAL) | | 2MASS photometry flag |
| (SQF_J) | | GLIMPSE photometry flag |
| (SQF_H) | | GLIMPSE photometry flag |
| (SQF_K) | | GLIMPSE photometry flag |
| (SQF_3P6UM) | | GLIMPSE photometry flag |
| (SQF_4P5UM) | | GLIMPSE photometry flag |
| (SQF_4P8UM) | | GLIMPSE photometry flag |
| (SQF_8P0UM) | | GLIMPSE photometry flag |
| (AP_LS_FLG) | | "Local Spitzer" photometry flag (Kuhn et al. 2013b) |
| (ORIGIN_J) | | origin of photometry |
| (ORIGIN_H) | | origin of photometry |
| (ORIGIN_K) | | origin of photometry |
| (ORIGIN_3p6um) | | origin of photometry |

Table 7—Continued

| Column Label (1) | Units (2) | Description (3) |
|---|---|---|
| (ORIGIN_4p5um) | | origin of photometry |
| (ORIGIN_5p8um) | | origin of photometry |
| (ORIGIN_8p0um) | | origin of photometry |
| **SED Properties** (Povich et al. 2013) | | |
| (SED_FLG) | | classification from SED analysis |
| (SED_AV) | mag | Av from SED analysis |
| (SED_STAGE) | | YSO stage |
| **X-ray Classification** (Section 3) | | |
| (H1_prior) | | class prior probability (position-dependent) |
| (H2_prior) | | class prior probability (position-dependent) |
| (H3_prior) | | class prior probability (position-dependent) |
| (H4_prior) | | class prior probability (position-dependent) |
| (H1_posterior) | | class posterior probability |
| (H2_posterior) | | class posterior probability |
| (H3_posterior) | | class posterior probability |
| (H4_posterior) | | class posterior probability |
| (H2_dominant_factor) | | dominant classification term[b] |
| Assign | | classification (0=unclassified, 1=H1, 2=H2, 3=H3, 4=H4) |

[a]X-ray source labels identify a *Chandra* pointing; they do not convey membership in astrophysical clusters.

[b]*H2_dominant_factor* reports the classifier term that exerts the most influence on the H2 posterior probability (1 = prior, 2 = MedianEnergy, 3 = $J$ magnitude, 4 = X-ray variability, 5 = spectral type, 6 = 4.5 $\mu$m magnitude, 7 = infrared SED model).

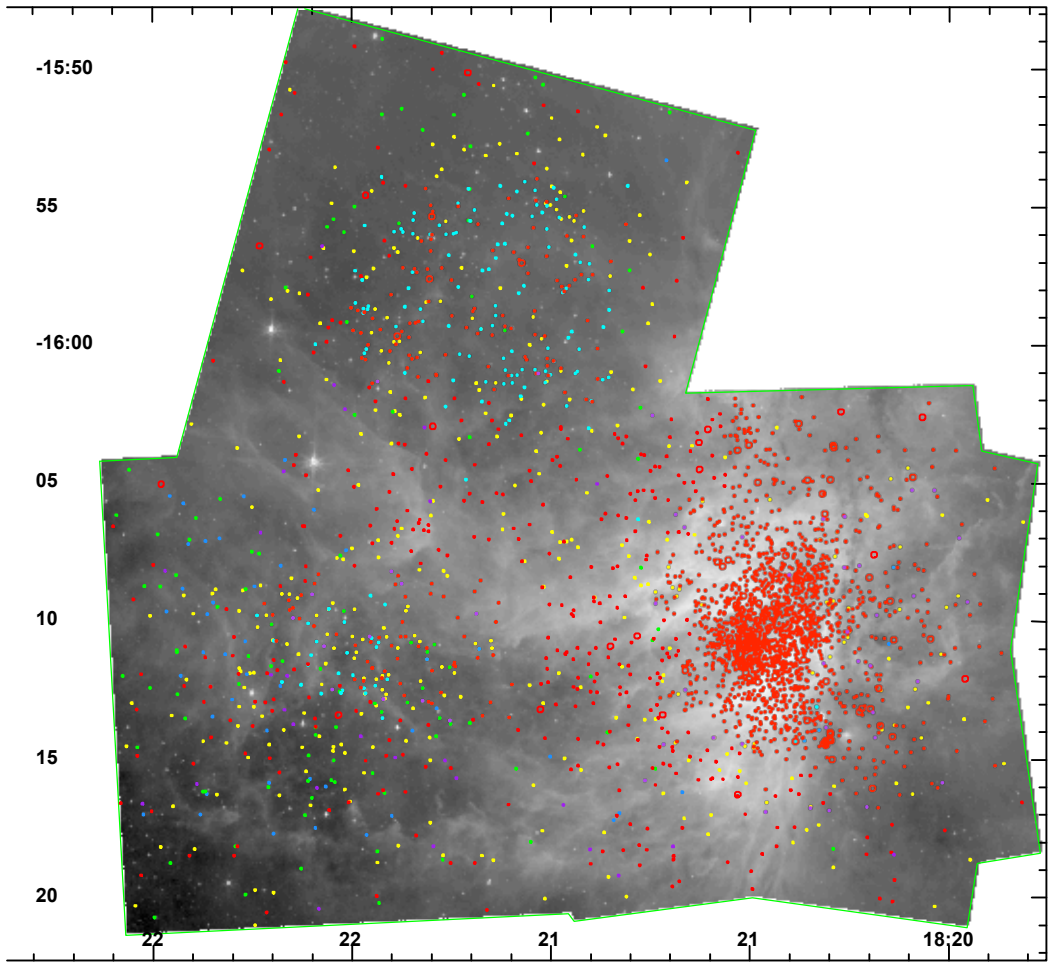

Fig. 4.—: Classes assigned to all M 17 X-ray sources, shown on an 8 $\mu$m *Spitzer* image: unclassified=yellow, H1=purple, H2=red (large circle indicates IR excess), H2-overturned (high H2 posterior but low evidence, Section 3.3) =cyan, H3=green, H4=blue. H2 sources are propagated to the MPCM catalog (S2); MPCMs that are not X-ray sources are *not* shown here.

## 5. Validation of X-ray Classification

Table 8 compares, for each MSFR, the number of X-ray contaminants predicted by simulations (columns 2,5,8,11) to the tallies of H1, H3, H4, and unclassified X-ray sources (columns 3,6,9,12). Some tallies are also expressed in parentheses as a percentage of the X-ray catalog.

A consistent pattern is clearly seen: the number of sources asserted to be contaminants (H1, H3, H4) represents only a small fraction of the contaminant populations predicted by simulations. We attribute this behavior in our classifier to the fact that the prior probabilities for each of our three contaminant classes are much lower than those of the member class, over most of the *Chandra* fields of view. For example, the "median(prior)" columns in Table 8 report the median H1, H3, and H4 class prior probabilities across the *Chandra* fields of view. Many classifiers exhibit low recovery rates for minority classes (e.g. Waske et al. 2009; Sug 2011).

We are encouraged to see that the total number of possible contaminants identified by the classifier (Col. 12) is in reasonable agreement with the number of the simulated contaminants (Col. 11). In most MYStIX fields, between 20% and 30% of the X-ray sources are predicted to be, and are classified as, contaminants or unclassified sources. Note that the fraction of sources identified as "unclassified" is an uninteresting performance metric, because an observer can achieve any desired unclassified fraction via the design of the decision rule. A conservative decision rule that declares a classification only when there is an overwhelming victor among the posterior probabilities will produce a large unclassified fraction, whereas a liberal decision rule that assigns a classification to every source would of course achieve a zero unclassified fraction.

Table 8. Comparing Contaminant Simulation Predictions to Classifier Results

| MSFR | foreground | | | background | | | extragalactic | | | all contaminants | |
|---|---|---|---|---|---|---|---|---|---|---|---|
| | sim. | H1 | median(prior) | sim. | H3 | median(prior) | sim. | H4 | median(prior) | sim. | H1+H3+H4+Unclassified |
| (1) | (2) | (3) | (4) | (5) | (6) | (7) | (8) | (9) | (10) | (11) | (12) |
| Flame Nebula | 9 | 0 | 0.02 | 2 | 0 | 0.01 | 84 | 7 | 0.17 | 95 (17%) | 125 (23%) |
| RCW 36 | 16 | 9 | 0.04 | 16 | 0 | 0.04 | 86 | 0 | 0.18 | 118 (24%) | 165 (33%) |
| NGC 2264 | 48 | 0 | 0.04 | 18 | 0 | 0.02 | 221 | 126 | 0.18 | 287 (22%) | 430 (32%) |
| Rosette Nebula | 109 | 4 | 0.07 | 68 | 0 | 0.04 | 404 | 190 | 0.23 | 581 (30%) | 625 (32%) |
| Lagoon Nebula | 101 | 2 | 0.05 | 197 | 3 | 0.10 | 193 | 102 | 0.10 | 491 (20%) | 599 (25%) |
| NGC 2362 | 50 | 0 | 0.08 | 35 | 0 | 0.06 | 159 | 119 | 0.24 | 244 (35%) | 223 (32%) |
| DR 21 | 39 | 4 | 0.06 | 39 | 5 | 0.06 | 87 | 0 | 0.13 | 165 (22%) | 171 (22%) |
| RCW 38 | 63 | 13 | 0.09 | 25 | 1 | 0.04 | 124 | 1 | 0.18 | 212 (21%) | 206 (20%) |
| NGC 6334 | 71 | 8 | 0.06 | 63 | 0 | 0.05 | 99 | 2 | 0.08 | 233 (15%) | 125 ( 8%) |
| NGC 6357 | 109 | 11 | 0.06 | 141 | 0 | 0.07 | 168 | 13 | 0.09 | 418 (18%) | 408 (17%) |
| Eagle Nebula | 179 | 7 | 0.09 | 135 | 1 | 0.06 | 230 | 106 | 0.12 | 544 (19%) | 765 (27%) |
| M 17 | 240 | 73 | 0.11 | 221 | 107 | 0.12 | 270 | 47 | 0.13 | 731 (24%) | 703 (23%) |
| W 3 | 142 | 36 | 0.09 | 43 | 3 | 0.03 | 342 | 29 | 0.22 | 527 (25%) | 523 (25%) |
| W 4 | 43 | 3 | 0.07 | 18 | 0 | 0.03 | 117 | 71 | 0.18 | 178 (28%) | 236 (36%) |
| Trifid Nebula | 97 | 3 | 0.15 | 26 | 10 | 0.05 | 67 | 38 | 0.11 | 190 (30%) | 215 (34%) |
| NGC 3576 | 113 | 1 | 0.09 | 36 | 0 | 0.03 | 139 | 0 | 0.11 | 288 (19%) | 391 (26%) |
| NGC 1893 | 114 | 7 | 0.10 | 11 | 0 | 0.01 | 224 | 132 | 0.21 | 349 (24%) | 332 (23%) |

Note. — Cols. 2, 5, 8, and 11 tally the number of contaminants predicted by simulations. Cols. 3, 6, 9, and 12 tally classification outcomes. Cols. 4, 7, and 10 report median class prior probabilities across each *Chandra* field of view.

Another approach to validating the X-ray classification is to compare X-ray and infrared properties for the various classes. The scatter plots of $J$-band magnitude versus X-ray flux shown in Figure 5, stratified by class assigned, provide several lines of support for the class assignments made to our X-ray sources.

**Pre-main Sequence (PMS) Populations** It is well known that the X-ray luminosities of PMS and main-sequence (MS) stars are correlated with their stellar masses, with the X-ray luminosities of PMS stars being elevated by a factor of ∼1000 above main sequence levels (Preibisch et al. 2005, Figure 7). The $J$-band magnitude and the apparent X-ray photon flux are good empirical surrogates for stellar mass and X-ray luminosity, respectively. A $J$-band magnitude versus X-ray flux correlation is thus naturally expected for either of the two classes of stars (PMS or MS) providing that these stars span a wide range in mass and are at a similar distance from us. In Figure 5 a clear $J$ vs. X-ray flux correlation is seen for the H2 class, as expected, and not seen for the other classes. The clearest correlations are found for the MSFRs with rich, lightly-absorbed PMS populations that have been captured with deep NIR and X-ray exposures, such as NGC 2362, NGC 2264, Lagoon Nebula, Rosette Nebula, Eagle Nebula, and NGC 1893. For the MSFRs with rich but heavily-absorbed PMS populations (DR 21, NGC 6334, and M 17), absorption lowers the apparent $J$-band flux of many H2 objects (seen as a scatter of points upward from the main H2 locus).

**Extragalactic sources** Several MSFRs (Rosette Nebula, NGC 2362, and NGC 1893) lie away from the Galactic plane, have deep NIR/X-ray observations, and have patchy and/or shallow molecular clouds. In those figure panels the locus of the brightest extragalactic sources (H4; at $J > 18$ mag) has been detected and is distinct from the H2 locus.

**Foreground stars** $J$-band magnitude histograms for the simulated populations of foreground stars often have two peaks, a dimmer peak from numerous M-type field dwarfs and a brighter peak from other types of field stars (Figure 3). These two foreground peaks often overlap with the single but wider peak in the MSFR member (H2) distribution. The dimmer foreground peak is often below the completeness limits of the NIR observations (purple dotted curve in Figure 3). These two factors typically lead to more frequent identification of individual brighter foreground stars, while dimmer foreground stars either remain unclassified or are incorrectly classified as MSFR members. The trend of increasing $J$-band magnitude, from H1 (black) to Unclassified/H2 (yellow/red) supports this notion, for example in the RCW 38, Eagle Nebula, M 17, W 3, and NGC 1893 panels.

**Possible Foreground (PFGD) Candidates** PFGD candidates can be identified independently from the MYStIX classifier, for example, by employing color cuts on a NIR color-color diagram combined with a cut in X-ray median energy (e.g., Kuhn et al. 2010; Getman et al. 2012). In Figure 5, PFGD sources are marked (cyan circles) when $J-H < 0.65$ mag and $ME < 1.2$ keV. Since the NIR colors of field MS stars and lightly absorbed PMS stars are nearly indistinguishable (Getman et al. 2012, Figure 8a), a color selection alone is meaningless for the MSFRs with lightly absorbed PMS populations. Figure 5 shows that the color selection incorrectly flags as PFGD (cyan) hundreds of PMS stars (red) in the lightly obscured MSFRs NGC 2264, NGC 2362, and part of the Rosette Nebula. On the other hand, the color selection could be efficient for the regions with heavily absorbed PMS stars, for which the NIR colors of MS and PMS stars are different due to the reddening effect (Kuhn et al. 2010, Figure 3). Figure 5 shows that the classifier's results for Unclassified/H1 (yellow/black) are consistent with the color selection of PFGD (cyan) for the heavily absorbed MSFRs in the Rosette Molecular Cloud, DR 21, RCW 38, NGC 6334, NGC 6357, M 17, and W 3.

For PMS stars, a correlation between 3.6 $\mu$m magnitude and X-ray flux is expected. This correlation should be generally more scattered than the $J$-band magnitude versus X-ray flux correlation (Getman et al. 2012, Figure 4b), because [3.6] emission is boosted for the subset of stars that have dusty disks. Both these effects are observed for the MYStIX H2 sources.

Finally, Kuhn et al. (2013c) show that the majority of the clusters identified in the MYStIX MPCM lists are consistent with the clusters identified in previous optical/IR/X-ray studies, and show that many small and often previously unknown clusters are found to lie projected against known molecular cloud cores.

## 6. Limitations

Feigelson et al. (2013, Appendix B) discuss several limitations of the MYStIX data and analysis methods. We discuss below some technical issues that potentially limit the effectiveness of our X-ray source classification.

### 6.1. Astrophysical Limitations of the Contaminant Simulations

Recall from Section 3.2 and Appendix A that astrophysical simulations of foreground stars, background stars, and extragalactic objects play a central role in constructing the elements of the classification model. Several astrophysical issues can potentially impact the fidelity of these simulations. First, the Galactic population synthesis model we use (Robin et al. 2003) assumes uniform extinction throughout the Galactic plane and does not model spiral arms. Second, we do not model hard X-ray sources in the Galactic plane attributed to cataclysmic variables and other classes of accretion-driven X-ray binary systems (Hong et al. 2009). Third, we cannot verify the fidelity of the absorption maps (e.g., Figure 6a) that we constructed from the dust reddening of NIR field stars. These maps affect the apparent flux and thus the detectability of simulated background stars and extragalactic objects.

### 6.2. Uncertain Class Prior Probabilities

Recall from Section 3.1 that our class prior probabilities are calculated using density maps for detected sources and density maps for sources expected from the three contaminant classes. These maps necessarily suffer from Poisson noise arising from the finite source samples available (particularly the H2 training set). We regulate that noise by an adaptive smoothing process, but smoothed density estimates inevitably broaden sharp features found in the parent population (i.e., clusters are broadened by smoothing).

Even if we could perfectly estimate the spatial distribution of the contaminant classes, the simulations may not produce the correct normalizations, i.e., the total numbers of H1, H3, and H4 sources may be incorrectly predicted. Uncertainties in those predicted tallies of H1/H3/H4 sources lead directly to uncertainties in our class prior probabilities.

Standard Bayesian models do not allow for the prior distributions of the model parameters to themselves be uncertain. However, multi-level (or hierarchical) Bayesian models address this issue (Loredo 2012a,b; Congdon 2010). In a multi-level model for our classification problem, the true numbers of H1/H3/H4 contaminant sources in our catalog would themselves be cast as uncertain model parameters, with prior distributions estimated from the simulations. In other words, the simulations would produce not just a best-estimate of the H1/H3/H4 tallies, but a plausible distribution for each. As always, the Bayesian machinery would produce a joint posterior distribution for all the model parameters—in this case a four-dimensional distribution for source class, H1 tally, H3 tally, and H4 tally. When inferences about the *classification of individual sources* are desired (which is our goal in MYStIX), one would marginalize (integrate) over the contaminant tally parameters to obtain a one-dimensional posterior distribution for the source class parameter.

An interesting side effect of such a model is that it would enable inferences about the *class populations* without classifying individual sources. For example, if the true number of foreground stars in our X-ray catalog was of scientific interest, then one would marginalize (integrate) over the source class, H3 tally, and H4 tally parameters to obtain a one-dimensional posterior distribution for the H1 tally parameter.

### 6.3. The Likelihood Tail Problem

The standard formulation of Naive Bayes inference assumes that the PDFs for observed data, conditioned on the class ("likelihood functions", Section 3.2), are known perfectly. In our application, the four PDFs for median X-ray energy and the four PDFs for $J$ magnitude are smoothed density estimates obtained from finite samples of those source properties. Simulations provide the samples for the H1, H3, and H4 classes; the training set (Section 3.2.7) provides the sample for the H2 class. Such PDFs necessarily suffer from statistical uncertainties, and those uncertainties rise in the tails of the distributions where fewer and fewer data points are available.

For most sources, these uncertain tails have little effect on the class posterior probabilities. For example, consider in Figure 2 a source with a median X-ray energy of 2.5 keV. That measurement would produce an H1 likelihood of zero (ruling out H1), very small and uncertain likelihoods for H2 and H4, and a much larger and more certain likelihood for H3. Although the likelihood ratio between H2 and H4 is uncertain, the dominance of the H3 likelihood is clear.

However, if a measurement is rare for all classes, then all likelihoods are produced from uncertain tails. The behavior of the classifier will depend unstably on the details of the density estimation procedure and on the source samples used to infer the PDFs. Two examples of this are astrophysically interesting. The first example occurs at the hard end of the median X-ray energy distributions (Figure 2), e.g., above $\sim 4.5$ keV; sources this hard are very rare in all classes. Protostars in the MSFR are expected to lie here, and we should expect that the classifier will have difficulty distinguishing them from contaminants. Human astronomers also have difficulty distinguishing protostars from extragalactic sources using X-ray data.

The second example occurs at the bright end of the $J$ magnitude distributions (Figure 3), e.g., brighter than $\sim 8$ mag; sources this bright in $J$ are very rare in all classes. Massive stars are expected to lie here, and we should expect that the classifier will have difficulty distinguishing them from contaminants. Human astronomers also have difficulty distinguishing massive cluster members from foreground stars using $J$ magnitude.

In future studies, we hope to learn methods for incorporating likelihood uncertainty into Bayesian classification; that complexity was not possible within the resources of MYStIX.

### 6.4. Other Limitations

The H2 training set (Section 3.2.7)—required to estimate H2 likelihood functions for median X-ray energy, $J$ magnitude, and X-ray variability—may be biased with respect to the true member population detected by the X-ray observation. For example, since clustering and IR excess are among the few classifier terms available when the training set is defined, members that are clustered and/or exhibit an IR excess may be over-represented.

Likelihood functions are constructed from samples of source properties taken over the fields of view defined by the *Chandra* pointings in hand, not over astrophysically relevant fields of view. When such a field contains multiple populations with astrophysically distinct properties (e.g., age, absorption) a single classification model may not be optimal for any one population.

Since most members of the extragalactic class lie beyond the completeness limits of our NIR observations, the absence of an identified NIR counterpart clearly represents some degree of evidence for the H4 class. However, interpreting that evidence quantitatively—as four class likelihoods—would be a challenge, requiring models of the (spatially varying) NIR completeness limits and models of the performance of the X-ray/NIR matching algorithm for each class. As mentioned in Section 3.1, we instead handle missing data by omitting that likelihood term.

## 7. Summary

Except in rare cases such as the Orion Nebula Cluster, the identification of individual stellar members of a star-forming region has never been an easy task. Historically, one can see periods when one method dominated another, primarily based on available observational technologies. During the 1950s, repeated visual-band photographic exposures found variable members of nearby star-forming regions. This method is likely to see a resurgence with wide-field multi-epoch surveys of the Galactic Plane with solid state detectors, such as the Via Lactea project of VISTA, all-sky surveys like the All Sky Automated Survey and Catalina Real-time Transit Survey, and the planned LSST. During the 1960-80s and continuing today, H$\alpha$ emission from low resolution grism spectra provided an efficient way to locate accreting PMS stars. Also during this period, $UBV$ photometry located blue-colored stars that were spectroscopically confirmed as massive OB members. Starting in the 1980s and continuing today, infrared imagery (particularly from satellite-borne telescopes) provided excellent samples of young stars with dusty disks, revealed by their greatly enhanced emission in mid-infrared bands. During the 1990-2000s, X-ray telescopes identified many PMS stars and OB stars in rich young clusters.

Each method has its limitations. Infrared excess populations are restricted to younger PMS stars that still harbor disks. H$\alpha$ is mostly restricted to a subclass of disk-bearing stars where disk gas is accreting onto the star. Both visual-band and infrared observations are often hampered by bright nebulosity from the H II regions around clusters and by dust obscuration from surrounding molecular clouds. With current X-ray instrumentation and at distances typical for MYStIX MSFRs, detecting young stars below $1 M_\mathcal{S}$ requires an extraordinary investment in observing time, such as that devoted to the Orion Nebula Cluster (Getman et al. 2005a).

Each method generates false positives—stars or galaxies in the field of view that are falsely identified as young stars. Many classes of old stars exhibit variability in the visual-band. Foreground dM3 stars are H$\alpha$ emitters. Asymptotic giant branch post-main sequence stars can have dusty envelopes. Extragalactic sources can be both faint X-ray sources and faint infrared-excess sources.

In the MYStIX project we tackle this challenge by combining X-ray, near-infrared, and mid-infrared data from modern telescopes (*Chandra*, 2MASS, UKIRT, and *Spitzer*) to give an answer to the question: Which detected objects are members of the star-forming region? By combining lists of X-ray detected members (Section 3), IR-detected members (Povich et al. 2013), and OB stars we have constructed a catalog of 31,784 "MYStIX Probable Complex Members" in 20 MSFRs, which are available electronically. A validation of the procedure for NGC 2264, where the stellar population had been extensively studied by other researchers over half a century, is described by Feigelson et al. (2013).

For X-ray detected sources, we use the probabilistic approach known as "Naive Bayes Classification", which is perhaps the simplest method of multivariate classification from the machine learning community. This framework provides a coherent method for combining observational evidence that carries classification information. This machinery for combining evidence comes at the cost of constructing a statistical model of the observations—estimating the odds of finding each class in the X-ray catalog (the class priors) and the PDF of each observable quantity for each class (the class likelihood functions).

## A. Contaminants in MYStIX X-ray Catalogs

### A.1. Modeling Galactic and Extragalactic Contaminant Populations

X-ray surveys of star-forming regions suffer contamination by extragalactic sources, mainly quasars and other active galactic nuclei, which can be seen even through the Galactic plane as faint, absorbed X-ray sources. For any star-forming region, additional contamination arises from foreground and background Galactic stars, mainly main-sequence stars and some types of giants. For regions located in the quadrant of the Galactic plane centered on the Galactic center, contamination by cataclysmic variables might also be important (Getman et al. 2011, and references therein).

We perform detailed simulations for extragalactic and Galactic X-ray contaminating populations expected in the direction of the MYStIX MSFRs. The methodology for such simulations is described in detail by Getman et al. (2011). The simulations take into consideration a variety of factors involving a Galactic population synthesis model (Robin et al. 2003), stellar X-ray luminosity functions, X-ray flux functions for extragalactic sources, *Chandra* telescope response, source detection methodology, and possible spatial variations in the X-ray background and absorption through molecular clouds.

Two major differences from Getman et al. (2011) pertain to the MYStIX contamination simulations: an improved source detection technique, and a different estimate of the absorption through molecular clouds (see Appendix A.2). In the previous work, removal of very weak simulated contaminants that would have fallen below the source detection threshold of corresponding real *Chandra* observations was based on a signal-to-noise criterion. For MYStIX, simulated contaminants were "detected" using criteria more consistent with the source detection process we apply to *Chandra* observations (Broos et al. 2011a).

Basic properties of the simulated contaminating populations—spatial distributions, X-ray median energies, and J-band magnitudes—are then employed in the "Naive Bayes" X-ray source classifier to establish membership probabilities for each of the MYStIX X-ray sources (Section 3.2). Figure 6 shows that the spatial distributions of the simulated contaminants may well reflect the effects of the inhomogeneous X-ray exposure and cloud absorption across the field. These effects, along with the simulated X-ray median energies and J-band magnitudes, are important discriminants for distinguishing different populations of MYStIX sources.

## A.2. Absorption Maps for MYStIX Fields

As shown in Figure 6(a), modeling distant contaminant populations requires a map of absorption associated with the molecular cloud in each MYStIX complex. Maps of molecular gas line emission were not available for many targets at the time the simulations were performed. We thus constructed approximate maps of absorption by dust through the molecular clouds using the reddening of background stars (cf. Lombardi & Alves 2001; Dobashi 2011; Schneider et al. 2011). A variety of approximations are made to construct the $A_V$ maps, but the maps give a rough sense of absorption by the clouds in these star-forming regions. The method described below produced $A_V$ maps for all MYStIX regions except W 3, where $A_V$ was derived from the $^{12}$CO 2-1 map of Bieging & Peters (2011) scaled to the values of our W4 North region $A_V$ map, and the Flame Nebula, where $A_V$ was derived from the $C^{18}O$ map from (Aoyama et al. 2001) and the gas-to-dust relationship from Ryter (1996).

We use the MYStIX infrared source catalogs (2MASS, UKIRT, *Spitzer*/IRAC) to estimate reddening on different lines of sight through the cloud. The $A_V$ maps are primarily based on the deepest available NIR data. However, we start to run out of NIR detection at absorptions $> 10$ mag in the $V$ band, so regions with high absorptions and low star counts are supplemented by MIR sources. We assume the reddening law of Rieke & Lebofsky (1985) for $JHK$ and a combination of Flaherty et al. (2007) and Rieke & Lebofsky (1985) for [3.6][4.5]. NIR intrinsic stellar colors were taken from Lombardi & Alves (2001), and we assume $[3.6 - 4.5]_0 = 0$.

Since our goal is to estimate absorption to background stars (behind the molecular cloud), we try to remove foreground stars and complex members from the IR catalogs. Likely foreground stars are identified by $A_V$ values that are lower than expected for the distance to the cluster, assuming an average absorption of 0.7 mag kpc$^{-1}$. Possible disk-free cluster members are identified by X-ray counterparts, and disk-bearing cluster members are identified by their infrared excess (Simon et al. 2007). Some foreground stars and complex members will be missed by this screening, and may bias the $A_V$ estimates.

For each target, $A_V$ estimates to thousands of individual stars are transformed into a raw $A_V$ map using kernel smoothing, with a kernel size adapted for the IR source density of the region. For each target, $A_V$ is also estimated for stars within an annular control region outside the cloud. A tilted plane model of the control region absorptions is subtracted from the raw $A_V$ map. The resulting maps of the molecular cloud's absorption are shown in Figure 7. The spatial structure of the clouds in our maps mostly match *Herschel* 500 $\mu$m images shown in Kuhn et al. (2013c). For $A_V < 10$ our maps are similar to those presented by Dobashi (2011). Comparisons to $CO$ studies from the literature typically show agreement to better than a factor of 2 for the densest regions (see Kuhn et al. 2013c, and references therein).

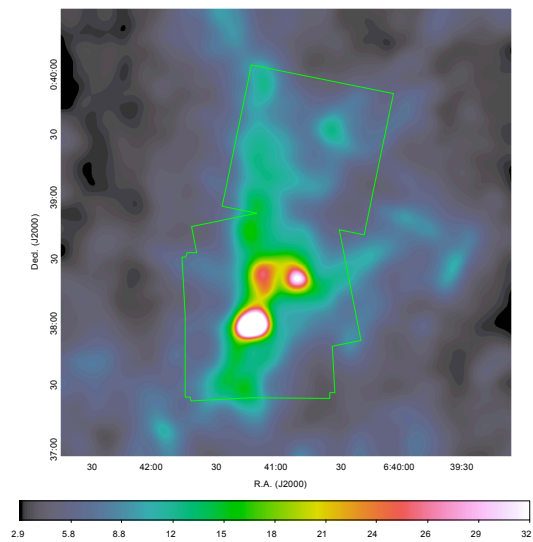

Fig. 7.—: Visual-band absorption map for NGC 2264, shown in units of magnitude ($A_V$). Corresponding figures for each MSFR listed in Table 8 are available in the electronic edition of this article.

Acknowledgments: The MYStIX project is supported at Penn State by NASA grant NNX09AC74G, NSF grant AST-0908038, and the *Chandra* ACIS Team contract SV4-74018 (G. Garmire & L. Townsley, Principal Investigators), issued by the *Chandra* X-ray Center, which is operated by the Smithsonian Astrophysical Observatory for and on behalf of NASA under contract NAS8-03060. M. S. Povich was supported by an NSF Astronomy and Astrophysics Postdoctoral Fellowship under award AST-0901646. We thank Steve Majewski and Remy Indebetouw for access to results from the *Spitzer* Vela-Carina survey. This research made use of data products from the *Chandra* Data Archive and the *Spitzer Space Telescope*, which is operated by the Jet Propulsion Laboratory (California Institute of Technology) under a contract with NASA. This research used data products from the United Kingdom Infrared Telescope (UKIRT), which is operated by the Joint Astronomy Centre on behalf of the Science and Technology Facilities Council of the U.K.; some UKIRT data were obtained as part of the UKIRT Infrared Deep Sky Survey (Lawrence et al. 2007) and some were obtained via UKIRT director's discretionary time. This research used data products from the Two Micron All Sky Survey, which is a joint project of the University of Massachusetts and the Infrared Processing and Analysis Center/California Institute of Technology, funded by the National Aeronautics and Space Administration and the National Science Foundation. The HAWK-I near-infrared observations were collected with the High Acuity Wide-field K-band Imager instrument on the ESO 8-meter Very Large Telescope at Paranal Observatory, Chile, under ESO programme 60.A-9284(K). This research has also made use of NASA's Astrophysics Data System Bibliographic Services, the SIMBAD database operated at the Centre de Données Astronomique de Strasbourg, and SAOImage DS9 software developed by Smithsonian Astrophysical Observatory.

*Facilities:* CXO (ACIS), Spitzer (IRAC), FLWO:2MASS (), CTIO:2MASS (), UKIRT (WFCAM)

– 46 –Kissler-Patig, M., et al. 2008, A&A, 491, 941

Kuhn, M. A., et al., MYStIX X-ray paper

Kuhn, M. A., et al., MYStIX mid-IR paper

Kuhn, M. A., et al., MYStIX ellipsoid cluster modeling paper

Kuhn, M. A., Getman, K. V., Feigelson, E. D., et al. 2010, ApJ, 725, 2485

Lawrence, A., Warren, S. J., Almaini, O., et al. 2007, MNRAS, 379, 1599

Lombardi, M., & Alves, J. 2001, A&A, 377, 1023

Loredo, T. J. 2012, in "Statistical Challenges in Modern Astronomy V," (Lecture Notes in Statistics, Vol. 209), ed. Eric D. Feigelson and G. Jogesh Babu; arXiv:1206.4278

Loredo, T. J. 2012, in "Statistical Challenges in Modern Astronomy V," (Lecture Notes in Statistics, Vol. 209), ed. Eric D. Feigelson and G. Jogesh Babu; arXiv:1208.3035

Megeath, S. T., Gutermuth, R., Muzerolle, J., et al. 2012, AJ, 144, 192

Ochsenbein, F., Bauer, P., & Marcout, J. 2000, A&AS, 143, 23

Paolillo, M., Schreier, E. J., Giacconi, R., Koekemoer, A. M., & Grogin, N. A. 2004, ApJ, 611, 93

Povich, M. S., Churchwell, E., Bieging, J. H., et al. 2009, ApJ, 696, 1278

Povich, M. S., Smith, N., Majewski, S. R., et al. 2011, ApJS, 194, 14

Povich, M. S., et al., MYStIX SED paper

Preibisch, T., Hodgkin, S., Irwin, M., et al. 2011, ApJS, 194, 10

Preibisch, T., Kim, Y.-C., Favata, F., et al. 2005, ApJS, 160, 401

Naylor, T., et al., MYStIX source macthing paper

Rieke, G. H., & Lebofsky, M. J. 1985, ApJ, 288, 618

Robin, A. C., Reylé, C., Derrière, S., & Picaud, S. 2003, A&A, 409, 523

Ryter, C. E. 1996, Ap&SS, 236, 285

Schneider, N., Bontemps, S., Simon, R., et al. 2011, A&A, 529, A1

Shemmer, O., Brandt, W. N., Vignali, C., Schneider, D. P., Fan, X., Richards, G. T., & Strauss, M. A. 2005, ApJ, 630, 729

Simon, J. D., Bolatto, A. D., Whitney, B. A., et al. 2007, ApJ, 669, 327

Skiff, B. A. 2009, VizieR Online Data Catalog, 1, 2023

Stelzer, B., Flaccomio, E., Montmerle, T., et al. 2005, ApJS, 160, 557

Sug, H. 2011, *Intl. J. Mathematical Models and Methods in Applied Sciences*, 5, 797

Telleschi, A., Güdel, M., Briggs, K. R., Audard, M., & Palla, F. 2007, A&A, 468, 425

Townsley, L. K., Broos, P. S., Corcoran, M. F., et al. 2011, ApJS, 194, 1